\documentclass[sigconf, nonacm]{acmart}

\newcommand\vldbdoi{XX.XX/XXX.XX}
\newcommand\vldbpages{XXX-XXX}
\newcommand\vldbvolume{14}
\newcommand\vldbissue{1}
\newcommand\vldbyear{2020}
\newcommand\vldbauthors{\authors}
\newcommand\vldbtitle{\shorttitle} 
\newcommand\vldbavailabilityurl{https://github.com/namanhboi/rdma_anns}
\newcommand\vldbpagestyle{plain} 

\usepackage[linesnumbered,ruled,vlined]{algorithm2e}
\usepackage{svg}
\usepackage{float}

\usepackage{xcolor}
\usepackage{todonotes}

\newcommand{\name}{\NoCaseChange{BatANN}}
\newcommand{\bigann}{BIGANN}
\newcommand{\texttwoimage}{Text2Image}
\newcommand{\msspacev}{MSSPACEV}
\newcommand{\deep}{DEEP}

\begin{document}
\title{Passing the Baton: High Throughput Distributed Disk-Based Vector Search with \name{}}

\author{Nam Anh Dang}
\affiliation{%
  \institution{Cornell University}
  \city{Ithaca}
  \state{New York}
}
\email{nd433@cornell.edu}

\author{Ben Landrum}
\orcid{0009-0003-4557-8850}
\affiliation{%
  \institution{Cornell Tech}
  \city{New York}
  \state{New York}
}
\email{blandrum@cs.cornell.edu}

\author{Ken Birman}
\orcid{0000-0003-2400-149X}
\affiliation{%
  \institution{Cornell University}
  \city{Ithaca}
  \state{New York}
}
\email{ken@cs.cornell.edu}

\begin{abstract}


Vector search underpins modern information-retrieval systems, including retrieval-augmented generation (RAG) pipelines and search engines over unstructured text and images. As datasets scale to billions of vectors, disk-based vector search has emerged as a practical solution. However, looking to the future, we must anticipate datasets too large for any single server and throughput demands that exceed the limits of locally attached SSDs. We present \name{}, a distributed disk-based approximate nearest neighbor (ANN) system that retains the logarithmic search efficiency of a single global graph while achieving near-linear throughput scaling in the number of servers. Our core innovation is that when accessing a neighborhood which is stored on another machine, we send the full state of the query to the other machine to continue executing there for improved locality. On 1B-point datasets at 0.95 recall using 10 servers, \name{} achieves \textbf{3.5--5.59$\times$} the throughput of the natural scatter--gather baseline and \textbf{1.44-2.09$\times$} the throughput of DistributedANN, while maintaining mean latency below \textbf{3 ms}.  Moreover, these results are achieved on a commodity network with standard TCP. To our knowledge, \name{} is the first open-source distributed disk-based vector search system to operate over a single global graph.

\end{abstract}

\maketitle

\pagestyle{\vldbpagestyle}
\begingroup\small\noindent\raggedright\textbf{PVLDB Reference Format:}\\
\vldbauthors. \vldbtitle. PVLDB, \vldbvolume(\vldbissue): \vldbpages, \vldbyear.\\
\href{https://doi.org/\vldbdoi}{doi:\vldbdoi}
\endgroup
\begingroup
\renewcommand\thefootnote{}\footnote{\noindent
This work is licensed under the Creative Commons BY-NC-ND 4.0 International License. Visit \url{https://creativecommons.org/licenses/by-nc-nd/4.0/} to view a copy of this license. For any use beyond those covered by this license, obtain permission by emailing \href{mailto:info@vldb.org}{info@vldb.org}. Copyright is held by the owner/author(s). Publication rights licensed to the VLDB Endowment. \\
\raggedright Proceedings of the VLDB Endowment, Vol. \vldbvolume, No. \vldbissue\ %
ISSN 2150-8097. \\
\href{https://doi.org/\vldbdoi}{doi:\vldbdoi} \\
}\addtocounter{footnote}{-1}\endgroup

\ifdefempty{\vldbavailabilityurl}{}{
\vspace{.3cm}
\begingroup\small\noindent\raggedright\textbf{PVLDB Artifact Availability:}\\
The source code, data, and/or other artifacts have been made available at \url{\vldbavailabilityurl}.
\endgroup
}

\section{Introduction}
\label{sec:intro}

In recent years, improvements in representation learning and the rise of retrieval augmented generation (RAG) \cite{lewis2020rag} as a technique to improve accuracy and reduce hallucinations in LLM outputs have driven a surge in interest in vector search. Typically, an embedding model is used to encode unstructured data, such as a chunk of a document \cite{adams_distributedann_2025, huang_learning_2013} or an image \cite{radford_learning_2021} as a high dimensional vector. After a collection of items has been encoded this way, the same model or one trained jointly with the embedding model can be used to represent queries in such a way that among the embeddings, the nearest neighbors of a query vector correspond to items in the dataset relevant to that query. 

The embedding process itself is inexact, hence finding the exact nearest neighbors of a query vector is neither practical nor necessary (known data structures for exact nearest neighbor search have query complexity which is either linear in the number of vectors or exponential in the dimensionality of the vectors \cite{chavez_searching_2001}). Instead, we use approximate nearest neighbor (ANN) search, finding a large fraction of the embeddings nearest to a query.


Because web-scale datasets can include billions of embeddings which are collectively too large to keep in memory, research into efficient disk-based search methods has been an active area. Today's state-of-the-art methods, such as DiskANN\cite{jayaram_subramanya_diskann_2019} and Starling\cite{wang_starling_2024} rely on search graphs, an index type that supports empirically logarithmic query time with respect to dataset size \cite{jayaram_subramanya_diskann_2019, yu_a_malkov_efficient_2020}. Considerable research has focused on optimizing the performance of these graph structures. However, the throughput of a single server running a disk-based system is bottlenecked by the aggregated bandwidth of a set of SSDs attached to a single machine. We show in \autoref{subsec:scalability} that even as you increase the number of search threads on a single server running a disk-based index, throughput remains stagnant as SSD bandwidth is completely saturated. This motivates \emph{distributed disk-based search}, which allows the system to serve more queries per second (QPS) by leveraging multiple machines in parallel. This is the context for the present paper, which starts by assuming that a large data set has been sharded (partitioned) across a cluster of compute nodes.

There are two natural ways to distribute a graph-based index. 
The first is to partition the dataset into disjoint subsets and build a graph index independently for each \cite{wang_milvus_2021, deng_pyramid_2019, gottesburen_unleashing_2024}. At query time, the system searches all partitions and merges their results. 
The second is to construct a single global graph over the entire dataset and then partition the nodes of the graph—i.e., its neighbor lists and embeddings—across multiple servers. 

Because the first method does not fully exploit the sub-linear scaling of search graphs, recent work has pushed toward scalable distributed traversal of a global ANN graph.  For example,  CoTra \cite{zhi_towards_2025} offers an in-memory distributed vector search system that leverages RDMA networking.  DistributedANN \cite{adams_distributedann_2025} proposes a disk-based distributed system built on commodity networking hardware. These systems differ in how they orchestrate inter-server communication during search, but both generally rely on a request-reply pattern to explore off-server neighbors during traversal.

We present \name{}\footnote{Pronounced ``baton'', a reference to the relay-like behavior of the query procedure.}, a new approach for distributed disk-based vector search over a global graph that achieves high throughput and low latency with standard TCP networking. The core innovation of our system is its asynchronous, state-passing query procedure, which forwards entire query states between servers. This design maximizes data locality and allows for all servers to stay busy while doing the minimal amount of work to advance a query.


We make the following contributions: 

\begin{itemize}
 \item We propose a new distributed search mechanism for disk-based vector search that achieves near linear scaling in throughput in high recall regimes even on TCP.
 \item We show that our system has minimal latency penalty when scaling up the number of servers and sustains its low latency even at very high query rates. 
 \item We open source our implementation and evaluate its performance against best available baselines in distributed disk based search on various 100M and 1B scale datasets.
\end{itemize}


\section{Background}
\label{sec:background}


Formally, $k$ nearest neighbor ($k$-NN) search is defined as follows. Given a set of vectors $X \subset \mathbb{R}^d$, a distance function $\delta:(\mathbb{R}^d \times\mathbb{R}^d) \rightarrow \mathbb{R}$ and a query vector $q \in \mathbb{R}^d$, find a set $\mathcal{K} \subseteq X$ such that $|\mathcal{K}| = k$, and $\max_{p \in \mathcal{K}} \delta(p, q) \leq \min_{p \in X \setminus \mathcal{K}} \delta(p, q)$. We use `point' and `vector' interchangeably to refer to elements of $X$. There exist other notions of nearest neighbor search (most notably $\epsilon$-NN), but we focus exclusively on $k$-NN as it's the dominant paradigm in the literature and practical application. In our experiments, $\delta$ is squared euclidean distance (L2) for the \bigann{}, \deep{}, \msspacev{} datasets, and for the \texttwoimage{} dataset, which uses Maximum Inner Product Search (MIPS), $\delta$ is the negative dot product between two vectors.

Graphs have been studied as a method of indexing vectors for nearest neighbor search since the late 1990s \cite{arya_approximate_1993}, although it is only recently that they have become so dominant for scalable, low-latency vector search \cite{jayaram_subramanya_diskann_2019, yu_a_malkov_efficient_2020, dongEfficientKnearestNeighbor2011}. In a \emph{search graph}, each vector in the dataset being indexed corresponds to a node in a graph. The edges are chosen in such a way that greedy traversal of the graph yields near neighbors of a query vector: starting from some node in the graph and comparing the query to its neighbors, the neighbor nearest the query is chosen as the next node to expand. In naive greedy search, this process is repeated until a point is reached which does not have any neighbors closer to the query than itself. Our goal is to find $k$ neighbors, hence we employ {\em beam search}.  This algorithm maintains a {\em beam} consisting of the best $L$ points seen so far, and at each step explores the neighborhood of the best unexplored point in the beam, replacing points in the beam when it finds new candidates closer to the query. This repeats until the beam converges to a set of points that all have been explored. Pseudocode for the beam search algorithm is presented in \autoref{alg:beam_search}.

\begin{algorithm}[h]
\caption{Beam Search}
\label{alg:beam_search}
\KwIn{$G$ = graph, $q$ = query vector, $W$ = I/O pipeline width}
\KwOut{$k$ nearest vectors to $q$}

$s \gets$ starting vector\;
$L \gets$ candidate pool length\;
candidate pool $P \gets \{\langle s \rangle\}$, explored pool $E \gets \emptyset$\;

\While{$P \not\subseteq E$}{
    $V \gets$ top-$W$ nearest vectors to $q$ in $P$, not in $E$\;
    Read $V$ from memory or disk\;
    $E.\mathrm{insert}(V)$\;
    \For{$nbr$ in $V.\mathrm{neighbors}$}{
        $P.\mathrm{insert}(\langle nbr, \delta(nbr, q) \rangle)$\;
    }
    $P \gets L$ nearest vectors to $q$ in $P$\;
}
\Return $k$ nearest vectors to $q$\;
\end{algorithm}



Practical vector search systems often combine indexing with \emph{quantization}: methods that store a lossy representation of the vectors in the dataset to enable approximate distance comparisons with a much smaller footprint. For systems requiring high accuracy, some number of results larger than $k$ are retrieved using the quantized vectors, and \emph{reranked} using exact distance comparisons computed with the original vectors, with the $k$ nearest neighbors returned being the true nearest neighbors within the group selected with the quantized vectors.  

The most popular quantization scheme is product quantization (PQ) \cite{jegou_product_2011}. PQ is used in our system for almost all distance comparisons. To encode a set of vectors, PQ first splits each vector into subspaces,  each of which consists of a slice of the dimensions of the vector space. Then, given a parameter $b$ representing the number of bits to store per subspace, $k$-means clustering with $k = 2^b$ is done to find a set of cluster centroids for each subspace, which are stored along with the quantized vectors. The compressed representation of each vector is then the cluster assignments of each of its subspaces, which for feasible values of $b$ is considerably smaller than the original float-valued vector. At query time, a vector can be reconstructed by concatenating the centroids to which it was assigned, and distances can be computed with respect to the reconstructed vector. Optimized implementations favor constructing a \emph{codebook} consisting of precomputed distances between each centroid and the corresponding subspace for a query \cite{andreQuickerADCUnlocking2021,johnsonBillionScaleSimilaritySearch2021}. Distances can then be computed by summing the contributions of the centroids to which a point has been assigned. 

The accuracy of ANN search is measured in terms of \emph{recall}, which is the average fraction of points in an output of length $k$ which are at least as close to the query as the true $k$-th nearest neighbor.

\subsection{Disk-based vector search}

The problem of scaling fast and accurate vector search to dataset sizes beyond what can fit in DRAM on a single node has been an active area of research for several years. 
Our work builds on DiskANN, which was the first SSD-optimized solution able to index billions of vectors while preserving performance competitive with an in-memory index \cite{jayaram_subramanya_diskann_2019}. The index is designed around a paradigm in which in-memory PQ data is used to guide beam search to select the best $W$ candidate nodes in the beam to explore and read its full embedding and neighbor ID list from disk. All $W$ SSD reads are then issued in parallel.  DiskANN additionally introduces the Vamana \cite{jayaram_subramanya_diskann_2019} search-graph construction, which seeks to minimize the number of hops in the graph needed for search to converge, and proposes an efficient algorithm for constructing such a graph over the points in the dataset. Data within the graph is laid out so that each point's unquantized vector and neighborhood will fit within a 4KB disk sector. During beam search, a highly compressed PQ representation of the dataset is kept in memory and used to decide which nodes to search next.  

Additional prior work is reviewed in Section \ref{sec:unrelated}.

\section{Distributed Vector Search}
\label{sec:related}
Our work is best understood in the context of other implementations of distributed disk-based vector search.  These broadly fall into two categories, which we now review.

\subsection{Scatter--Gather Approaches}

\begin{figure}[h]
    \centering
    \includegraphics[width=\linewidth]{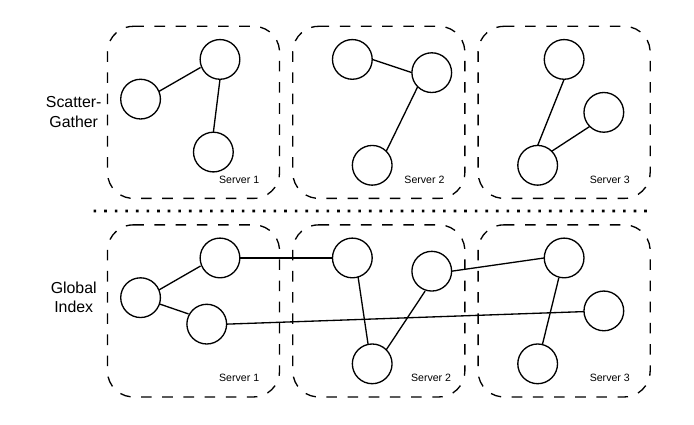}
    \caption{Scatter--Gather vs. Global Index}
    \label{fig:scatter_gather_global}
\end{figure}

\label{subsec:scatter-gather}

Many distributed vector search systems employ some variant of a scatter--gather paradigm.  As seen in Figure \ref{fig:scatter_gather_global}, these approaches split the dataset into shards which can be distributed to discrete nodes, and build indexing data structures on the points in each shard independently. At query time, a query vector is distributed to some or all of the shards (the ``scatter''), and each independently queries its local index for near neighbors. The results are then collected and merged (the ``gather and reduce'' step) into an approximation of the global near neighbors of the query. 

There is another variant of scatter--gather, which we refer to as scatter--gather Top-$N$, that relies on spatial partitioning in a manner similar to \name{}, and searches only the $N$ nearest partitions. We explore scatter--gather Top-$N$'s performance in \autoref{subsec:scatter_gather_top_n} and find that for high recall regimes, its throughput is at most equal to the traditional scatter--gather approach. Therefore, unless explicitly mentioned otherwise, we use ``scatter--gather'' to refer strictly to the traditional approach that searches all partitions.

Commercial vector databases using this approach such as  \cite{wang_milvus_2021, dilocker_weaviate_2025} are further discussed in \autoref{sec:prior-distributed}.

\subsection{Distributed Global Indices}

In contrast to the scatter--gather approach, DistributedANN \cite{adams_distributedann_2025} and CoTra \cite{zhi_towards_2025} build a global graph index over the full dataset, distribute it across several machines, and run queries on the global index using sophisticated data-dependent communication patterns which are meant to leverage the highly data-dependent behavior of queries on a graph index\footnote{Note that at the time of writing,  DistributedANN is not publicly available, so we re-implemented their system to compare against \name{}. CoTra is in-memory and uses RDMA, which makes it not a fair comparison against our system which is disk-based and uses TCP for communication.}. Provided that sufficient work can be done locally, the advantage of this global graph is its improved query complexity:  queries on search graphs are roughly logarithmic in the size of the dataset, and the sum of the logarithms of sizes of shards will always be larger than the logarithm of the sum \cite{adams_distributedann_2025}. This is the same broad approach used by our system. Because both systems are most relevant to our system, we choose to explain them more in-depth below:

\textbf{DistributedANN \cite{adams_distributedann_2025}} emulates DiskANN \cite{jayaram_subramanya_diskann_2019} in a distributed environment by replacing DiskANN's local SSD with a distributed key-value (KV) store. The system architecture is broadly divided into orchestration servers and scoring servers. An orchestration server maintains no state regarding the search graph; instead, it manages the state of query execution. Upon receiving a client query, it requests the initial search starting points from servers hosting the in-memory index.

Subsequently, the orchestration server advances the query state by dispatching distance comparison requests to the scoring servers, which contains a shard of the KV store with the search graph's data. These requests include the query's full embedding, its compressed representation, a threshold distance, and a target list of node IDs. For each node ID, the scoring server fetches the corresponding data from the KV store, calculates the full distance to the query, and computes the compressed distances to all of the node's neighbors. The threshold distance is used to determine whether the scoring server should include a neighbor's distance in the final result. Once these computations are complete, the scoring server returns the sorted full and compressed distances to the orchestration server. The orchestration server then updates its candidate set and full distance set before issuing the next batch of queries, effectively mirroring a beam search iteration in DiskANN.


Notice that {\em each step of beam search requires a round trip between the orchestrator and one or more servers with relevant vectors}.  Our work departs from DistributedANN in viewing these round-trips as problematic in servers that need to run at the highest possible query rates: each time an orchestrator thread pauses to collect results, resources are locked up on the orchestrator, and because the number of concurrent orchestrator threads running in a node is limited, we could reach a state in which nodes are underutilized because their orchestrator threads are all waiting.


\textbf{CoTra \cite{zhi_towards_2025}} is an in-memory distributed search system that builds a global graph which is distributed across nodes networked with remote direct memory access (RDMA) hardware. At query time, a central routing index identifies `primary' shards which contain a large number of relevant points, and `secondary' shards which are expected to be less relevant. Each primary node then maintains its own beam, and issues asynchronous requests and remote reads to other primary and secondary partitions to get neighborhoods and distance comparisons. The beams of the primary nodes are regularly synced in a procedure called co-search. 



Notice that both the above approaches rely on a send/receive pattern of communication where some query state on one server is dependent on distance computation results or one-sided RDMA reads of data held in the memories of other servers. In the next section, we discuss the \name{} architecture, which introduces a new asynchronous flow communication pattern in which entire queries are handed off from server to server as computation progresses.  The focus shifts: in \name{}, our goal will be to do as much work as we can before resorting to inter-server communication. 

\section{The \name{} Approach}

\begin{figure}[h]
    \centering
    \includegraphics[width=\linewidth]{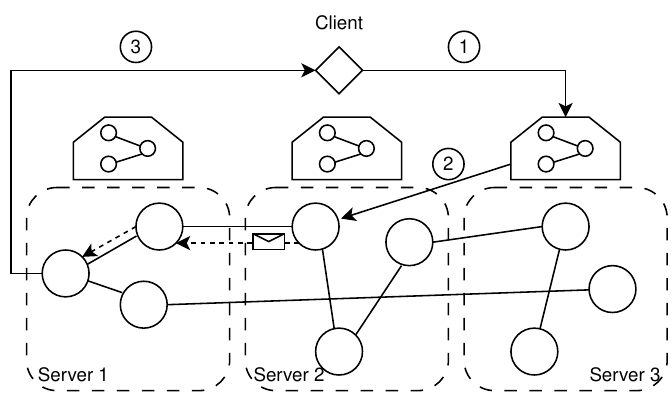}
    \caption{System overview}
    \label{fig:system}
\end{figure}

\label{sec:methods}

As discussed by Zhi et al., a single global index brings the benefit that query times are logarithmic in the size of the search graph \cite{zhi_towards_2025}. It is not evident that this would still be true for a distributed collection of individually indexed and queried partitions. Indeed, a problematic trend is evident in the experiments we will present in Section ~\ref{sec:experiments}: with existing scatter--gather approaches, the number of distance comparisons and disk I/O rises proportionally with the number of servers involved.





Our problem can thus be stated as follows: {\em Given a global search graph, what is the most efficient way to distribute and query it across many nodes without introducing unacceptably high latency from communication?}  As we have seen, one direction focuses on relatively costly hardware, such as RDMA \cite{zhi_towards_2025} or CXL \cite{jangCXLANNSSoftwareHardwareCollaborative2023} networking.  There has been some work aimed at leveraging  low-latency disaggregated memory technologies, but these remain somewhat exotic today. We target a more common case, where nodes are connected by commodity networking hardware and communicate using TCP, which can sustain very high throughput but is not ideal for round-trip interactions.  Accordingly, our design prioritizes reducing the amount of time queries spend waiting on communication, in much the same way that single-node disk-based indices minimize the number of round trips to disk. 

To this end, we first construct a graph over the entire dataset and partition it across servers using a  technique from Gottesburen et. al \cite{gottesburen_unleashing_2024}, ensuring that consecutive traversal steps are likely to access nodes stored on the same machine. A query is then dispatched to one of these servers to first conduct search on an in-memory head index to get the starting nodes of a beam-search execution (\textcircled{1} and \textcircled{2} in \autoref{fig:system}). During beam-search execution, when an off-server computation is required---i.e., when all current best candidate nodes reside on a remote server---we avoid the overhead of request–response round-trips by \textbf{sending the full query state} directly to the destination server (the envelope in \autoref{fig:system}). This distributed beam search then executes until all nodes in the beam are explored, at which point the last active server sends the result back to the client (\textcircled{3} in \autoref{fig:system}).


In this highly asynchronous paradigm, communication latency  is reduced compared to a traditional request-response model; as soon as the query state reaches another server, that server can take over execution using its local data.  The sender can garbage collect state that was previously in use for the query, freeing resources for use in subsequent queries. Unlike \name{}, which uses a single server to advance the state of a query at any given time, DistributedANN uses a request-response protocol to execute each hop across multiple scoring servers in parallel, and during each such stage the sender retains intermediate state. As we show in \autoref{subsec:multi_server_throughput}, this multi-server dependency reduces its overall throughput compared to \name{}. Furthermore, while parallelizing each hop across multiple servers can reduce latency despite the cost of an additional network hop to return results, we demonstrate in \autoref{subsec:latency} that \name{} achieves lower average latency than other systems simply by increasing its I/O pipeline width, $W$, to 64. Latency is also reduced with the help of neighborhood-aware graph partitioning. A partition/server will advance a query state for as many steps as possible until it has to transfer this state to another server.

By sending the entire state of the beam search execution for a query, our approach avoids performing additional distance computations compared to DiskANN on a single server for $W = 1$. 
For higher $W$, by using \autoref{alg:distributed-beam} to adapt the I/O pipeline to a distributed graph, we are also able to hold the number of distance comparisons done by \name{} to be approximately the same as for a single instance of DiskANN.




\subsection{What is a State?}

We can see from Algorithm~\ref{alg:beam_search} that beam search progresses step-by-step. In disk-based search, each step takes the $W$ closest frontier nodes in the beam—sorted by their approximate distances to the query—and issues disk reads for their full embeddings and neighbor IDs. Using the retrieved embeddings, we compute full-precision distances to the query, which will later be used for the final reranking stage. We then mark these $W$ nodes as explored and attempt to insert their neighbors into the beam based on in-memory PQ distances while maintaining a maximum beam size of $L$. The search concludes when all nodes in the beam have been explored, at which point we rerank the beam using the full-precision distances accumulated throughout the search to obtain the final $k$ results.

To advance a beam-search execution, the system must carry the beam state, and the full-precision result list, along with the parameters $L$, $k$, and $W$. For large beam sizes ($L \ge 200$), which are often necessary to achieve recall above 0.95, the total state size---including the query embedding---is approximately 4-8 KB. Transferring several kilobytes of state across machines for every inter-partition hop is non-trivial when using TCP. Consequently, minimizing the frequency of inter-server state transfers becomes critical for both throughput and latency. This motivates the next two components of the system: the in-memory head index and the global graph partitioning.

\subsection{In-Memory Head Index}
From Zhang et al. \cite{zhang_vbase_2023}, we know that a given run of beam search can generally be divided into two distinct phases. During the first phase, the algorithm approaches the general neighborhood of the query, and during the second phase, beam search stabilizes near the neighborhood of the query. This means that the beam explores many portions of the graph before settling into a region near the query. Since we are partitioning the graph across multiple servers, the first phase is likely to involve inter-server communication. Similar to CoTra and DistributedANN, we employ an in-memory head-index built from a 1\% sample of the graph to determine the starting nodes of beam-search. This head-index is replicated across all servers (\textcircled{1} \autoref{fig:system}).

\begin{figure}[h]
    \centering
    \includegraphics[width=\linewidth]{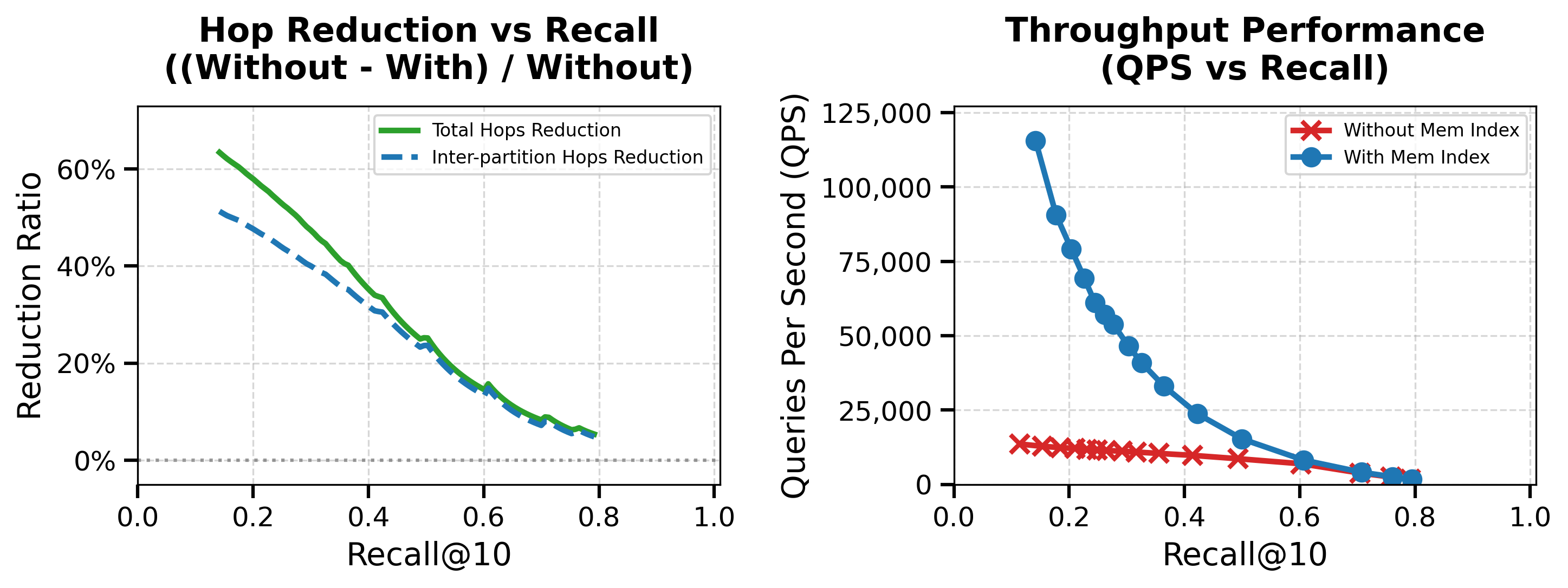}
    \caption{Comparison of number of hops and inter-partition hops vs recall and throughput vs recall for \texttwoimage{} 100M on 10 servers with and without mem index. As recall increases, in-mem index has less of an effect on throughput because its effect on hop count diminishes.}
    \label{fig:mem_ablation}
\end{figure}

Although more sophisticated approaches exist for this initial routing stage, such as building a KD-tree during index construction to guide routing at query time \cite{xuTwostageRoutingOptimized2021}, we choose the 1\% sampling method for its simplicity and demonstrated effectiveness in prior disk-based vector search systems \cite{wang_starling_2024, adams_distributedann_2025}. \autoref{fig:mem_ablation} demonstrates the effectiveness of the in-memory index. At lower recall values, which correspond to a smaller candidate list size $L$, we observe a reduction of over 50\% in both total and inter-partition hops. $L$ correlates closely with the total number of hops. For smaller $L$ values (around 10–50), the in-memory index bypasses a larger proportion of the total hop count because the overall hop count is relatively low. Because each hop incurs both disk I/O and distance computation, we see a large increase in throughput for lower recall values when using the memory index. As recall increases, corresponding to $L$ values between 400 and 1600, the impact of the in-memory index on throughput diminishes. This occurs because the skipped hops constitute a progressively smaller fraction of the total hops required for the search.

\subsection{Graph Partitioning}
To reduce the number of times a state has to be sent between servers, we can partition the graph so that nearby points are likely to reside on the same server. As this is analogous to the property of search graphs that makes neighbors likely to be connected by an edge, spatial partitioning algorithms like balanced $k$-means \cite{ahaltCompetitiveLearningAlgorithms1990} can be used to put neighbors of a point on the same server, thereby leveraging compute-data locality. Balanced $k$-means is what CoTra \cite{zhi_towards_2025} uses to partition its global graph across servers. We use a partitioning algorithm called Graph Partitioning proposed by Gottesburen et al. \cite{gottesburen_unleashing_2024}, which is faster to run and is known to out-perform balanced $k$-means in the scatter--gather approach. Our intuition is that the algorithm is more effective in preserving spatial relationships between graph nodes, yielding better locality and reduced inter-server communication.
\begin{figure}[h]
    \centering
    \includegraphics[width=0.8\linewidth]{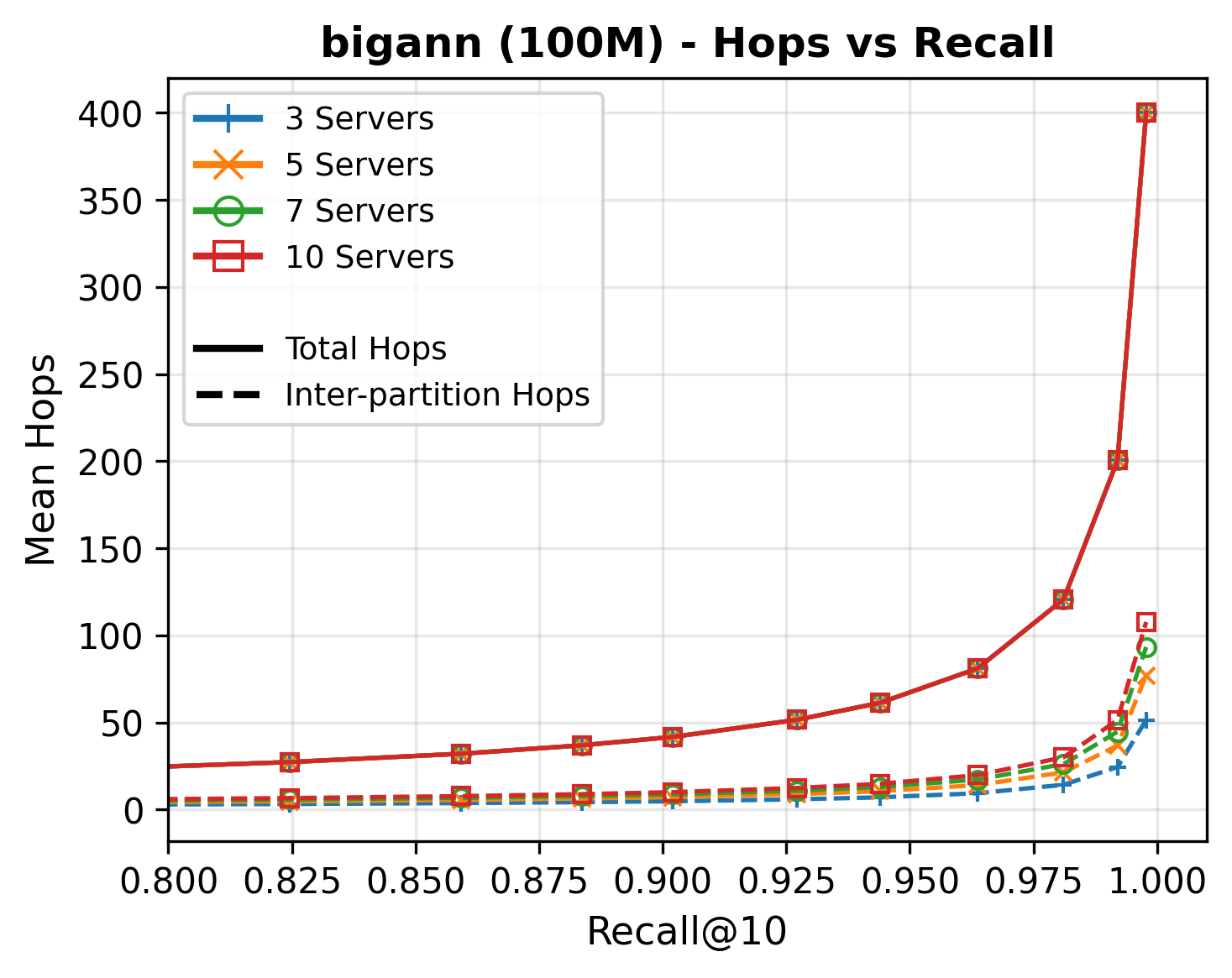}
    \caption{Number of hops vs. inter-partition hops for \bigann{} 100M on 3, 5, 7, 10 servers with $W = 1$. Inter-partition hops only account for 11.6-24.3\% of total hops, validating the effectiveness of graph partitioning.}
    \label{fig:hops_vs_inter_hops}
\end{figure}

To isolate the effect of graph partitioning, we will explore the number of inter-partition hops for best first search, a variant of beam search in which $W = 1$. With best-first search, at any given step, the beam will only choose to explore the best unexplored candidate node. From \autoref{fig:hops_vs_inter_hops}, we observe that at 0.95 recall@10, inter-partition hops account for 11.6\%, 17.34\%, 21.22\% and 24.3\% of total hops in the 3-, 5-, 7- and 10-server configurations, respectively. Because each inter-partition hop incurs communication and serialization overhead, these proportions translate directly to latency. We also see that the total number of hops is identical regardless of the number of servers used, which is expected for best-first search because both systems index the same global graph. The low frequency of inter-partition hops validates our graph partitioning approach.

Although graph partitioning and the in-memory index are effective at reducing inter-partition hops, they cannot eliminate the need to transfer state entirely. CoTra \cite{zhi_towards_2025} experimented with various partitioning methods—all of which achieved similar performance—and found that none were able to completely remove inter-server communication.

\subsection{I/O Pipeline Width}
\label{subsec:pipeline-width}
In this subsection, we explore the effects of higher I/O pipeline width $W$ on our system's performance and justify the usage of $W = 64$, which increases the amount of distance computation and I/O, but significantly reduces inter-server communication.

DiskANN shows that increasing the I/O pipeline width $W$ can reduce search latency. Modern consumer-grade SSDs can sustain more than 300K random reads per second, so issuing $W$ reads concurrently incurs roughly the same latency as issuing a single read \cite{jayaram_subramanya_diskann_2019}. By processing multiple items in the beam at once, an I/O pipeline also decreases the total number of beam-search hops. However, larger values of $W$ necessarily introduce some computational waste: many of the vectors and neighbor lists retrieved at higher pipeline widths are not close enough to the query to be inserted into the beam, even though their embeddings were read and neighbor distances were computed. DiskANN found that values of $W \in \{2, 4, 8\}$ do not waste compute and SSD resources, which is consistent with our observations.

In a distributed global graph, strictly following the standard beam-search procedure is not possible because some nodes selected for the I/O pipeline may reside on remote servers’ SSDs. To support the I/O pipeline in this setting, we introduce a simple heuristic described in \autoref{alg:distributed-beam}.

\begin{algorithm}[h]
\caption{Heuristic for I/O Pipeline in Distributed Global Graph}
\label{alg:distributed-beam}
\KwIn{beam $P$, pipeline width $W$}
\KwOut{next action for beam search}

$V$ :=  top $W$ unexplored nodes of $P$

\If{$V$ has nodes on current server}{
    explore all nodes on current server in $V$
}
\Else{
    send state to the server containing the top node in $V$\;
}
\end{algorithm}


\begin{figure}[h]
    \centering
    \includegraphics[width=0.8\linewidth]{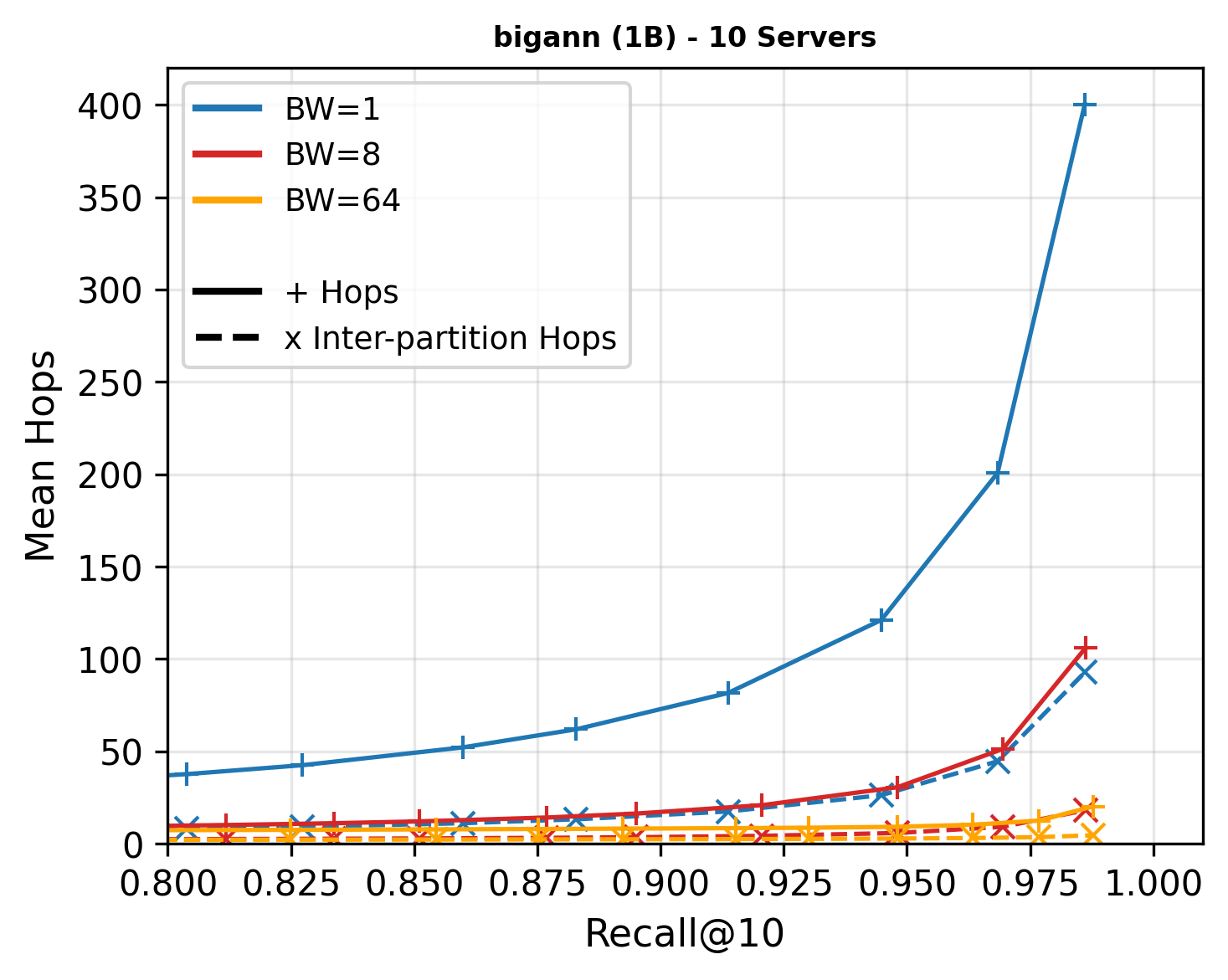}
    \caption{Comparison of inter-partition hops between $W = 1, 8, 64$ for \bigann{} 1B on 10 servers. For higher values of $W$, fewer total and inter-partition hops are needed.}
    \label{fig:inter_hop_bw_comparison}
\end{figure}


This heuristic serves two purposes. First, when all pipeline nodes are local, we benefit from the same latency improvements provided by the I/O pipeline in single-server settings. Second, it enables the I/O pipeline to operate efficiently in a distributed setting for larger values of $W$ by reducing the number of inter-partition hops, thereby lowering the overhead introduced by serialization and the network stack.

\autoref{fig:inter_hop_bw_comparison} shows that the total hop count decreases by a factor of five when increasing $W$ from 1 to 8, and almost a factor of four again from 8 to 64 at 0.95 recall@10. For $W = 1$ at 0.95 recall@10, the mean hop count is 138.6, for $W = 8$ it is 32.5, and for $W=64$ it is 9.3. This follows naturally from the fact that processing more nodes per iteration reduces the number of beam-search rounds/hops. Inter-partition hops scale proportionally with the total hop count for both $W$ values, accounting for 21.9\% of hops at $W = 1$, 18.8\% at $W = 8$, and 30.5\% at $W=64$. At $W=64$, the mean number of inter-partition hops is just around 2.8, meaning very little inter-server communication needs to happen.


\begin{figure}[h]
    \centering
    \includegraphics[width=\linewidth]{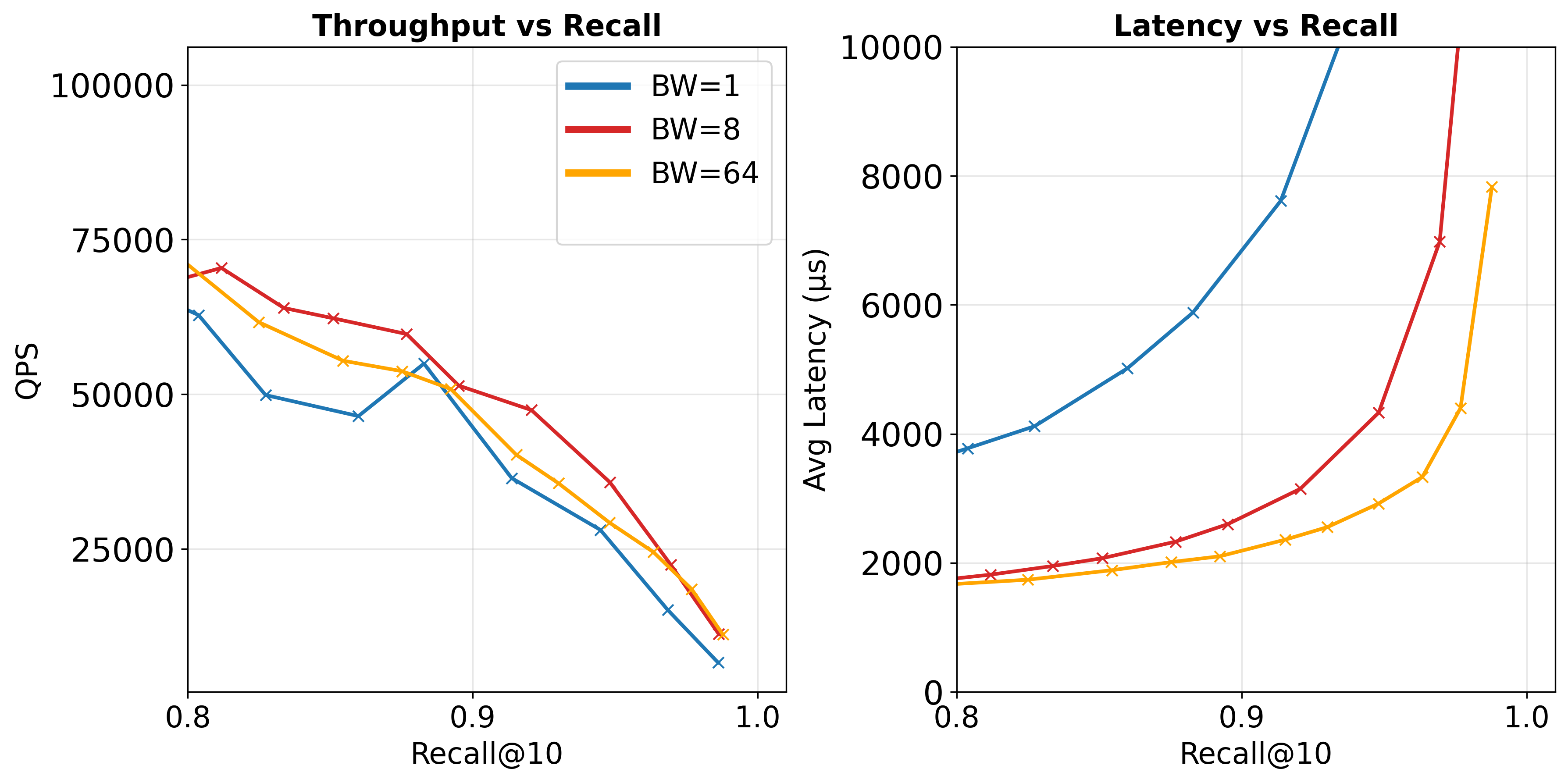}
    \caption{Throughput and latency comparison for $W = 1, 8, 64$ for \bigann{} 1B on 10 servers. $W=8$ has highest throughput, while $W=64$ has lowest latency.}
    \label{fig:throughput_latency_beam_width_comparison}
\end{figure}

We measured the number of distance computations and Disk I/O at $W=1,8,64$ and saw that these metrics are almost identical for both $W =1$ and $W= 8$ across all recall values. At $W=64$, there is an increase in both computation and I/O, especially for lower recall values. Distance computations and disk I/O are the 2 main bottlenecks for single-server disk-based vector search, while inter-server communication is a bottleneck for distributed vector search. 

From \autoref{fig:throughput_latency_beam_width_comparison}, we can see that $W=64$ strikes a balance between the throughput of $W=1, 8$, while being the clear winner in latency especially at higher recall values because of its low inter-server communication. We choose $W=64$ as the I/O pipeline width for \name{}.  At this value, \name{}'s throughput and latency consistently outperform those of ScatterGather and DistributedANN (see \autoref{sec:experiments}).




\section{Implementation and Optimizations}
\label{sec:implementation}
We implemented \name{} using C++ and base our single-server implementation on PipeANN's publicly available codebase \cite{guoAchievingLowLatencyGraphBased2025}. \texttt{io\_uring} was used for asynchronous I/O, and inter-server communication was handled by the ZeroMQ library \cite{zeromq_lib}, specifically the \texttt{PEER} socket. We also used the \texttt{moodycamel::ConcurrentQueue} \cite{desrochers_fast_2014} concurrent queue throughout the system. 

\begin{figure}[h]
    \centering
    \includegraphics[width=\linewidth]{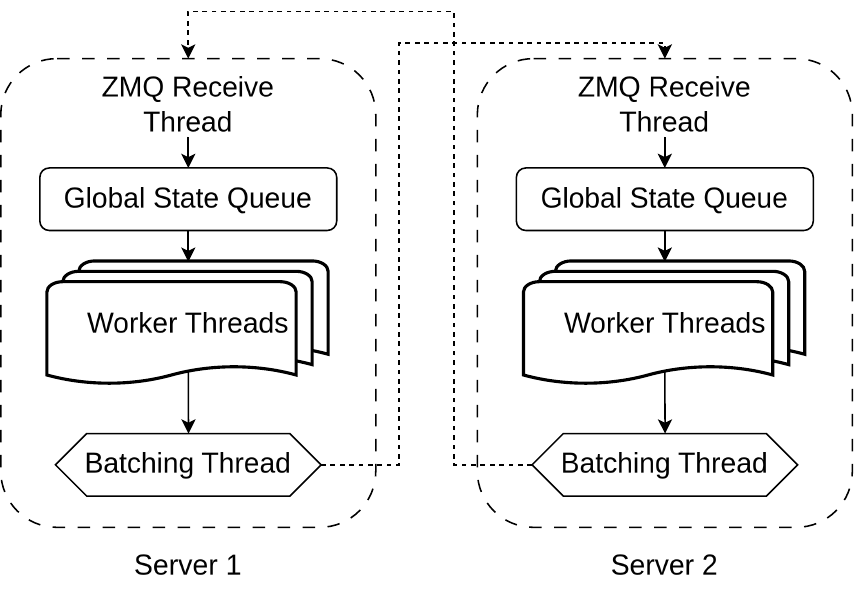}
    \caption{System implementation}
    \label{fig:system_impl}
\end{figure}

The inner structure of a single \name{} process is shown in \autoref{fig:system_impl}. The system runs a dedicated ZeroMQ receiver thread responsible for listening on a bound address, receiving incoming objects, and deserializing them before placing them into the appropriate queues. One such object is a query state, which the receiver places into a global state queue for the worker threads to consume. To send states or results, worker threads enqueue them into an outgoing queue that the batching thread monitors. The batching thread then opportunistically batches messages and sends them to other servers or clients. Below we present some optimizations found during system implementation.

\textbf{Inter-query balancing on a single thread:} In prior works on disk-based vector search \cite{jayaram_subramanya_diskann_2019, wang_starling_2024, tatsuno_aisaq_2025}, each search thread executes one query at a time. As observed by \cite{guoAchievingLowLatencyGraphBased2025}, this is suboptimal: during greedy search, each step requires reading the neighbor IDs of a candidate node from disk, leaving the compute thread idle while waiting for disk I/O to complete. This naturally motivates balancing multiple queries per thread, allowing the system to overlap computation for one query with asynchronous disk I/O (e.g., via \texttt{io\_uring}) for others. We build on PipeANN’s implementation of inter-query balancing \cite{guoAchievingLowLatencyGraphBased2025}, which they call CoroSearch. Their inter-query balancing approach processes queries in batches: once a thread is finished with a batch of queries, it tries to dequeue the next batch of queries/states and processes them concurrently. In contrast, our system maintains a fixed number of active queries (or states) per thread at all times. When a query finishes, the thread immediately pulls the next query from the queue, ensuring that the number of queries being balanced is fixed. We found that balancing 8 queries concurrently yielded the best overall performance. We discuss the throughput of CoroSearch, DiskANN and our approach in \autoref{subsec:single_server}.

\textbf{Caching query embeddings:} Query embeddings are sent to each server only once. A server caches the embedding until it receives an acknowledgment from the client indicating that the final result has been delivered. This avoids repeatedly transmitting the same embedding alongside every state message. This optimization is also used in CoTra \cite{zhi_towards_2025}.

\textbf{Pre-allocating objects}: To eliminate allocation overhead during message deserialization, we use object pooling for all message types exchanged between servers (e.g., search states and query embeddings). When a communication handler thread receives an incoming message, it deserializes directly into a pre-allocated object obtained from a lock-free queue (\texttt{moodycamel::ConcurrentQueue}\cite{desrochers_fast_2014}). This removes dynamic allocation from the critical receive path and improves system throughput. After processing, worker threads then return the pre-allocated objects to their respective pools for reuse.

\textbf{Memory footprint: } We keep a PQ representation of the full dataset in memory on each node, with each vector compressed to 32 bytes. For 100M and 1B scale, this equates to 3.2 GBs and 32 GBs respectively. This PQ data is used to compute the approximate distances to neighbors of a node, whose IDs are fetched from disk. There are recent works on disk-based graph search that propose storing neighbors' quantized data alongside neighbor IDs on disk that has superior throughput and latency to DiskANN \cite{chenAlayaLaserEfficientIndex2026}. We identify this as a promising extension for future work and discuss it in \autoref{sec:unrelated}. Additionally, we keep the map of node\_ids to partition ids, which is used to send the state of the beam-search execution. The partition ids are stored as uint8, and for 100M and 1B scale, this accounts for 0.1 and 1GB respectively. For points residing on a given node, we also keep a mapping of node ids to disk sector ids to know where to read from. This type of mapping is used in disk-based methods where nodes are not stored contiguously by id on disk like \cite{wang_starling_2024}. With a billion data points, the size of the mapping is less than 1 GB.

\section{Experiments}
\label{sec:experiments}
\begin{table}[hb]
  \caption{Datasets used in the experiments}
  \label{tab:datasets}
  \centering
  \begin{tabular}{cccccc}
    \toprule
    Dataset & Scale & Dim & Type & Similarity & \# queries \\
    \midrule
    \bigann{}    & 1B/100M & 128 & uint8 & L2 & 10000 \\
    \msspacev{}  & 1B/100M & 100 & int8  & L2 &  29316 \\
    DEEP      & 100M    & 96  & float & L2 & 10000 \\
    \texttwoimage{}      & 100M    & 200  & float & MIPS & 30000 \\
    \bottomrule
  \end{tabular}
\end{table}
\textbf{Setup:} We conduct all experiments on a CloudLab \cite{duplyakin_design_2019} cluster of 11 c6620 nodes connected by 25Gb-Ethernet. The c6620 nodes each have a 28-core Intel Xeon Gold 5512U at 2.1GHz and 128GB ECC Memory (8x 16 GB 5600MT/s RDIMMs) and run Ubuntu 22.04 LTS.

\textbf{Datasets:} We evaluate our system using four publicly available datasets, \bigann{}, \deep{}, \msspacev{}, and \texttwoimage{} from the Big-ANN benchmarks \cite{simhadri2024resultsbigannneurips23}. All datasets are evaluated at 100M scale, with \bigann{} and \msspacev{} also evaluated at 1B scale. \bigann{} contains 128-dim uint8 SIFT descriptors and serves as a classic large-scale ANN benchmark. \msspacev{} provides 100-dimensional int8 embeddings from Microsoft’s SpaceV model for semantic search. The \msspacev{} dataset is skewed, meaning queried vectors are closer together making it a more difficult dataset \cite{chengCharacterizingDilemmaPerformance2024}. DEEP consists of 96-dimensional floating-point descriptors extracted from deep convolutional networks. The \texttwoimage{} dataset uses Maxixium Inner Product distance and is bi-modal, consisting of images embedded using the SeResNext-101 model and text queries embedded using a DSSM model. Its vectors have 200 dimensions with each dimension represented as a 4-byte float. Its bi-modality and high dimensionality makes it the hardest dataset we tested. Together, these datasets cover a diverse range of vector types, dimensionalities, and metrics, enabling a broad evaluation of system performance. A summary is given in \autoref{tab:datasets}.

\textbf{Graph Construction:} All of our indices at both the 100M and 1B scales are Vamana graphs \cite{jayaram_subramanya_diskann_2019} built with parameters $R = 64$, $L = 128$, and $\alpha = 1.2$ for L2 and $\alpha = 1.0$ for MIPS, as suggested by Manohar et al.\cite{manoharParlayANNScalableDeterministic2024}. We built all graphs with ParlayANN \cite{manoharParlayANNScalableDeterministic2024} and merged them with the base files to create disk indices. Scalable distributed search graph construction has been explored in the literature \cite{shiScalableOverloadAwareGraphBased2025a,zhi_towards_2025} and would be valuable for a large-scale production deployment, but was not necessary at the scale of our experiments.

\textbf{Graph partitioning:} We partitioned the dataset with the Graph Partitioning method from \cite{gottesburen_unleashing_2024}. We use another server with 96 cores and 1.5 TB of DRAM to partition the billion scale datasets, as the memory requirements of their implementation make it infeasible to do on a c6620 Cloudlab node. 

\textbf{Baselines}: We compare our system against the ScatterGather approach described in \autoref{subsec:scatter-gather} and DistributedANN \cite{adams_distributedann_2025}:

\emph{ScatterGather}: Partitions are assigned using the same method as \name{} from \cite{gottesburen_unleashing_2024}. Each server uses the inter-query balancing approach we described in \autoref{sec:implementation}. The partitions use the same graph construction parameters as the full dataset, as experiments and precedent from prior work \cite{manoharParlayANNScalableDeterministic2024} suggest that parameter choice for Vamana graph construction is not sensitive to the size of the dataset being indexed. In our experiments, we call this baseline ScatterGather. We found that $W=8$ achieves the best throughput and latency for ScatterGather.

\emph{DistributedANN}:  DistributedANN does not provide an open-source implementation, but does provide a highly detailed description of the algorithm and its implementation.  Guided by this description, we re-implemented DistributedANN within our system. The orchestration server is analogous to a server in our system, with the primary difference being that it issues scoring queries to the scoring servers rather than performing direct disk I/O and distance computations. The scoring server also resembles a server in \name{}, except that it advances the search state by only a single step, utilizing the node IDs provided by the orchestration server as the initial frontier, before returning results to the orchestration server. Each scoring server performs direct disk I/O and distance computations, and maintains in-memory PQ data for all nodes. 

However, our implementation departs from the original DistributedANN architecture in three ways. First, we store all PQ data in memory rather than on disk. Second, we co-locate the in-memory index on the scoring servers themselves instead of utilizing dedicated head-index servers. Third, instead of using the random partitioning described in the DistributedANN paper, we used the same spatial partitioning method as \name{}, which in our experiments yielded a slight improvement in throughput. All of these differences only served to improve the performance of DistributedANN, and are used in both \name{} and ScatterGather. Therefore, we believe that our evaluation is fair and provides a best-faith attempt at representing the DistributedANN system. For DistributedANN, we found that $W = 64$ yields the best latency and throughput.

Each \name{}, ScatterGather, and DistributedANN server runs 8 search/scoring threads. DistributedANN runs an additional orchestration server with 8 orchestration threads meaning it uses 8 more threads than both ScatterGather and \name{}. This orchestration process resides on the same server as the client process and is physically separated from all the scoring servers.

\textbf{Metrics:} We measure throughput using plots of queries-per-second (QPS) versus recall@10, a value typical for approximate nearest neighbor search systems \cite{aumuller_ann-benchmarks_2020} . We also ran experiments for $K=1,100$ in \autoref{subsec:vary_k}.

\subsection{Single-Server Throughput}
\label{subsec:single_server}
\begin{figure}[h]
    \centering
    \includegraphics[width=0.8\linewidth]{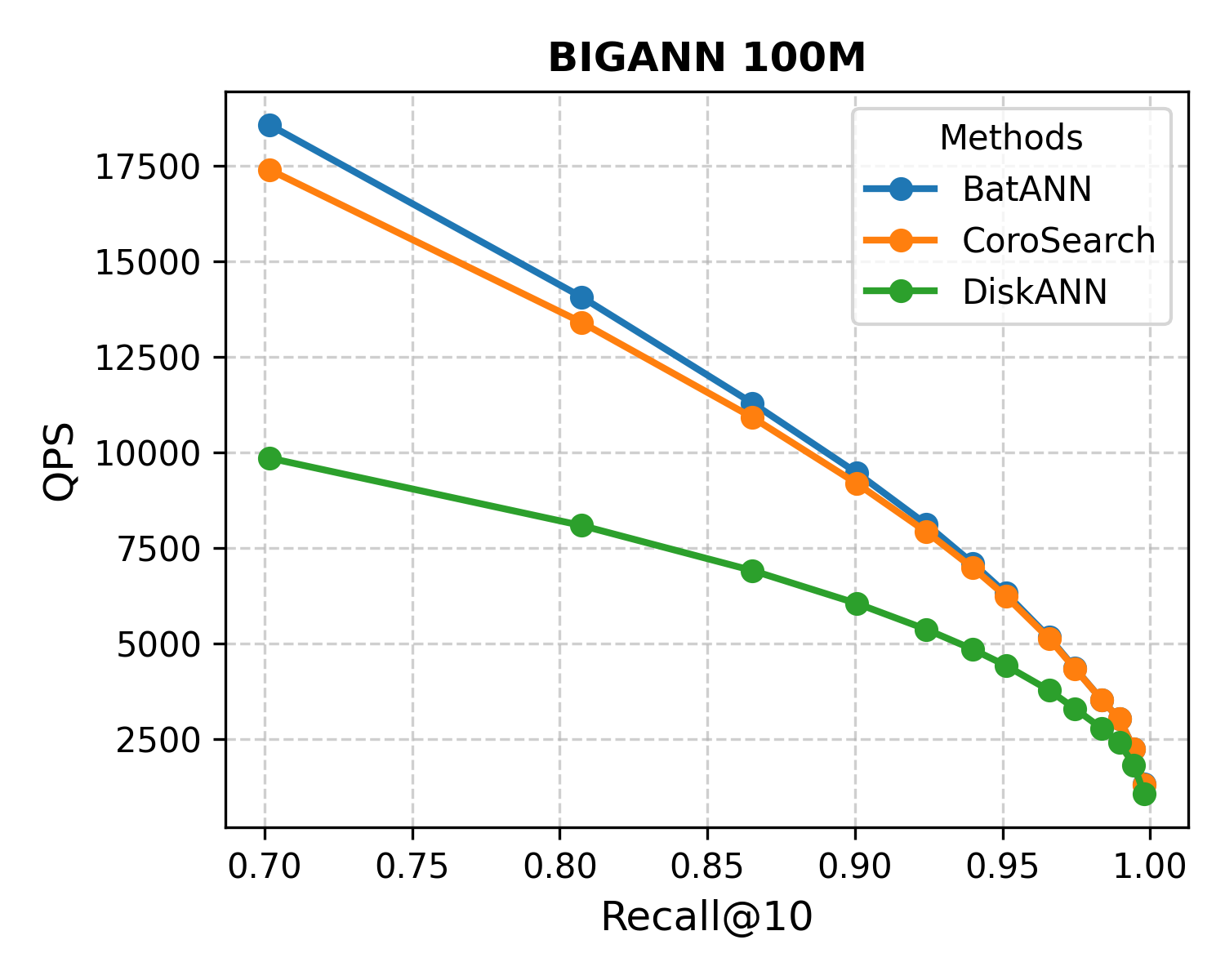}
    \caption{QPS recall graph of different search methods on \bigann{} 100M with 8 threads and $W = 64$}
    \label{fig:inter_query_balancing_bigann_100M}
\end{figure}
To measure the single-server throughput of \name{}, we issue queries from a client process on a separate server and utilize the full index on the search server. We re-implemented the queuing behaviors of DiskANN and CoroSearch within our system to compare them against our inter-query balancing approach, detailed in \autoref{sec:implementation}. As shown in \autoref{fig:inter_query_balancing_bigann_100M}, our approach provides consistently higher throughput across all recall regimes. Compared to DiskANN, which processes a single query at a time per thread, our approach continuously balances 8 queries or states concurrently on each thread. This effectively masks the substantial idle time that search threads otherwise spend waiting for disk I/O to complete. CoroSearch from PipeANN \cite{guoAchievingLowLatencyGraphBased2025} employs a similar strategy where each thread dequeues a batch of 8 queries, but it only dequeues another batch once the entire current batch is finished. Our method achieves slightly higher throughput than CoroSearch because it dynamically maintains 8 active states at all times, rather than waiting for a batch to finish completely before dequeing. For this reason, we adopt it as our single-server baseline for scaling experiments and as the underlying search procedure for our implementations of ScatterGather, \name{}, and DistributedANN.

\subsection{Multi-Server Throughput}
\label{subsec:multi_server_throughput}

\begin{figure*}[h]
    \centering
    \includegraphics[width=\textwidth]{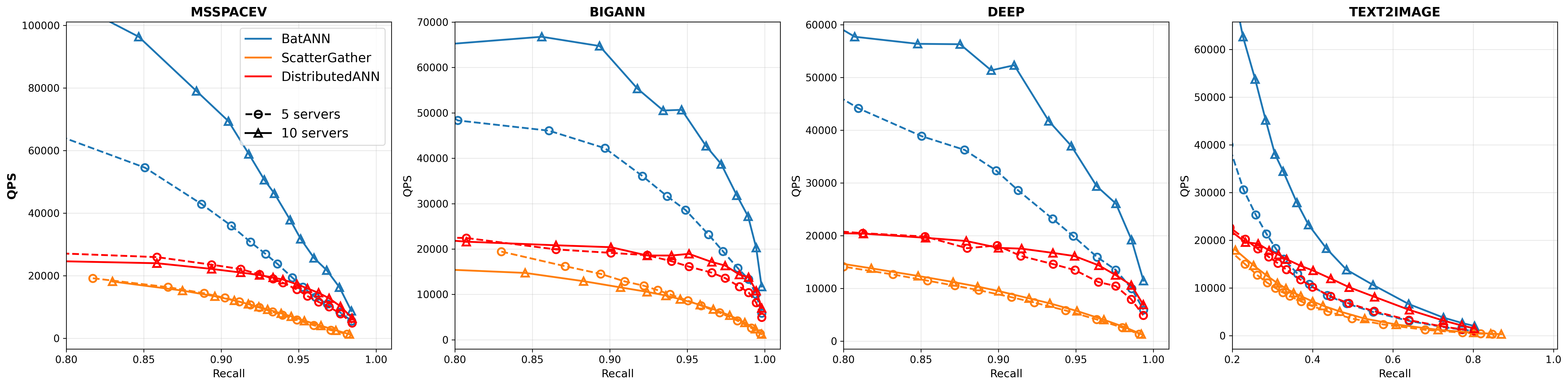}
    \caption{QPS recall curve of \bigann{}, \msspacev{}, \deep{} on 5 and 10 servers for 100M. \name{} outperforms ScatterGather on all setups. $W=64$ for DistributedANN and \name{}, $W=8$ for ScatterGather.}
    \label{fig:qps_recall_all_100M}
\end{figure*}

\begin{figure}[h]
    \centering
    \includegraphics[width=\linewidth]{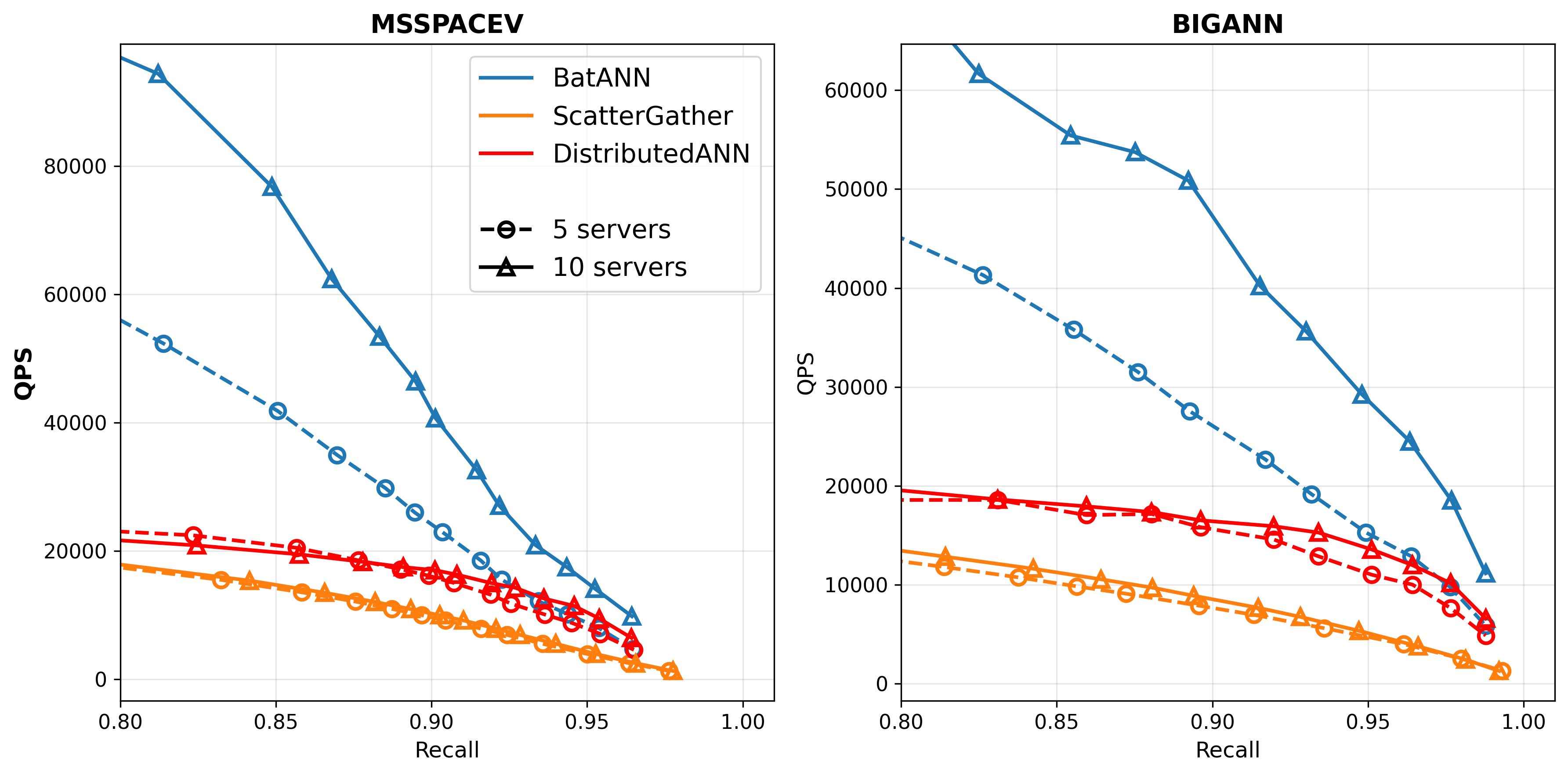}
    \caption{QPS recall curve of \bigann{}, \msspacev{} on 5 and 10 servers for 1B. \name{} outperforms ScatterGather on all setups with $W=64$ for \name{} and DistributedANN and $W=8$ for ScatterGather }
    \label{fig:qps_recall_1B}
\end{figure}





To measure throughput of all systems, we issue queries from a client process physically separate from all the search/scoring servers. For ScatterGather, we send each query to all search servers, and a query is finished once we receive a result from all search servers. For \name{}, we send the queries to the search servers in round-robin order. For DistributedANN, we issue all queries to the orchestration process living on the same machine as the client process. 

We calculate the throughput at 0.95 recall@10 for \bigann{}, \msspacev{}, and \deep{} and 0.7 recall@10 for \texttwoimage{}. \texttwoimage{} is a much harder dataset and no methods were able to achieve 0.9 recall@10 even for $L=1600$. We expect that the relative performance of all methods at 0.7 recall@10 is representative of higher recall regimes.

\textbf{Performance at 100M scale.} As can be seen in \autoref{fig:qps_recall_all_100M}, for 5 servers, \name{} achieves \textbf{2.98-3.70x} QPS improvement for the recall targets above across all datasets compared to ScatterGather, and \textbf{0.95-1.74x} compared to DistributedANN. For 10 servers, the improvement is even more dramatic. \name{} achieves \textbf{5.28-6.17x} QPS improvement compared to the ScatterGather baseline, and \name{} achieves \textbf{1.32-2.58x} improvement over DistributedANN. For both the 5 and 10 server setups, there is also broadly higher throughput than baselines across all recall regimes. \name{} only underperforms DistributedANN for 5 servers on \texttwoimage{}, where its throughput is 0.95x that of DistributedANN. We suspect that this is due to the fact that DistributedANN uses an additional 8 threads for its client-side orchestration server which are not available to \name{}.

\textbf{Performance at 1B scale.} From \autoref{fig:qps_recall_1B}, we see that \name{} continues to outperform ScatterGather and DistributedANN across all datasets and server counts. For 5 servers, \name{} throughput is \textbf{2.22-3.21x} that of ScatterGather, and \textbf{1.12-1.36x} that of DistributedANN. \name{} throughput improves for 10 servers, where its throughput is \textbf{3.50-5.59x} that of ScatterGather, and \textbf{1.44-2.09x} that of DistributedANN.

Overall, BatANN consistently outperforms and scales better from 5 to 10 servers than both ScatterGather and DistributedANN across all datasets. However, Figures 9 and 10 show that at high recall targets, BatANN's advantage over DistributedANN narrows, particularly on harder datasets like Text2Image and MSSPACEV. This occurs because high recall requires a larger candidate set size $L$, which increases the ratio of inter-partition hops. Since Algorithm 2 eagerly expands until the I/O pipeline is full of off-server nodes, larger $L$ values trigger elevated rates of eager expansion and inter-partition communication, progressively eroding BatANN's performance edge.


\subsection{Computation and I/O Efficiency}
\label{subsec:efficiency}

\begin{figure}[htp!]
    \centering
    \includegraphics[width=\linewidth]{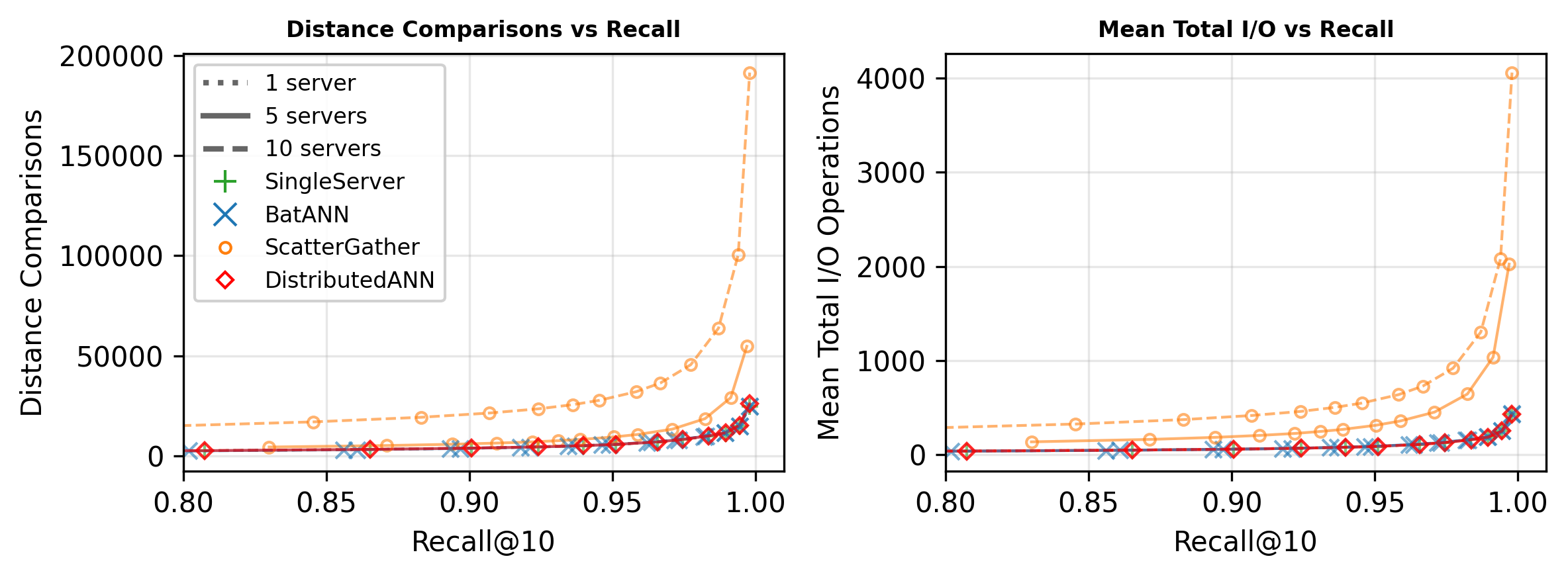}
    \caption{Distance comparisons and Mean of total I/O operations vs. recall of \bigann{} 100M on 5 and 10 server setups. \name{} and DistributedANN are clearly more computationally and I/O efficient} 
    \label{fig:dist_cmps_io_comparison}
\end{figure}


To measure computational efficiency we record the total number of distance comparisons performed per query, a standard method for measuring implementation-agnostic search graph performance \cite{manoharParlayANNScalableDeterministic2024}. 
To measure I/O efficiency, we record the mean number of disk I/O operations issued per query for each recall value. I/O and distance comparisons account for the vast majority of the runtime of a query.

For ScatterGather, each server independently searches its own shard, so the reported numbers are the sum of distance comparisons and disk I/O operations across all servers for each query. Consequently, the total amount of computation and disk I/O performed by ScatterGather increases proportionally with the number of servers, as shown in \autoref{fig:dist_cmps_io_comparison}.  In contrast, \name{} and DistributedANN use a global graph, which produces a clear I/O and computation advantage. Indeed, \name{} and DistributedANN perform nearly the same number of distance comparisons and I/O operations in both the 5- and 10-server configurations as they would on a single server. As a result, \name{} and DistributedANN scale much better than ScatterGather, as discussed in \autoref{subsec:scalability}. 


        

        



\subsection{Scalability}
\label{subsec:scalability}

\begin{figure}[htp!]
    \centering
    \includegraphics[width=\linewidth]{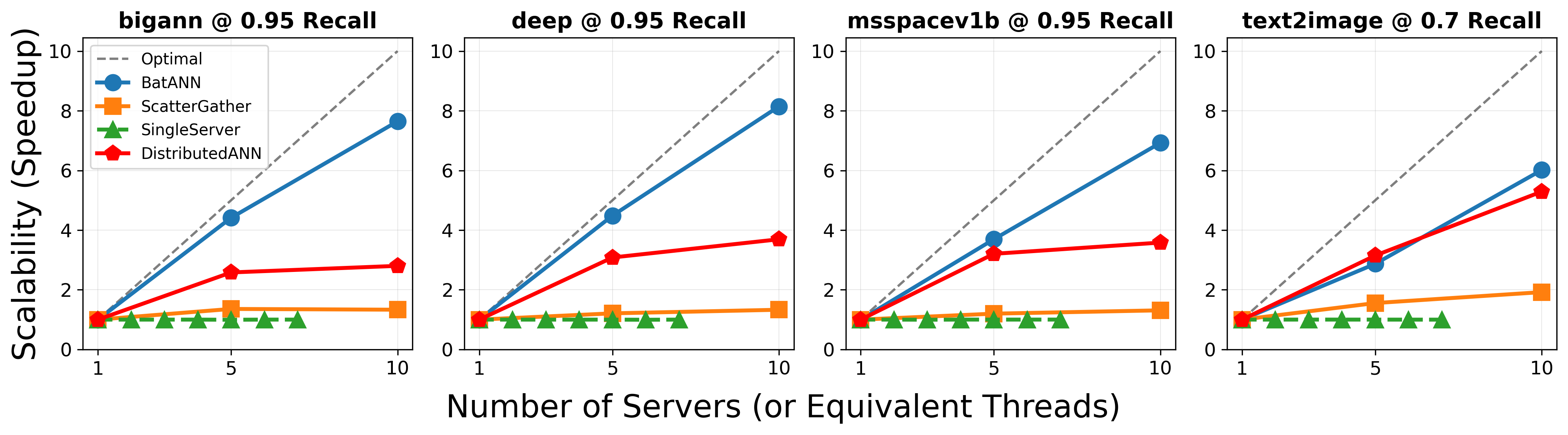}
\caption{Scalability on \bigann{}, \msspacev{}, and \deep{} for recall@10 = 0.95, and \texttwoimage{} for recall@10 = 0.70 at 100M up to 10 servers. \name{} scales very well across all datasets.}
\label{fig:all_scalability_100M}        
\end{figure}

To evaluate scalability, we measure the throughput of \name{}, DistributedANN, and ScatterGather on 5- and 10-server setups relative to a single-server baseline. Each search server uses 8 search threads, with DistributedANN requiring an additional 8 threads on a separate orchestration server. Speedup (\autoref{fig:all_scalability_100M}) is calculated by dividing the 5- and 10-server throughput by the throughput of a single server running 8 threads. As a further baseline, we also plot the single-server performance scaled from 16 to 56 threads. For all methods, queries are issued from a client machine.

From \autoref{fig:all_scalability_100M}, we see that \name{} benefits far more from strong scaling than ScatterGather. As shown by \autoref{fig:dist_cmps_io_comparison}, the number of distance comparisons and I/O operations remains functionally constant for our approach regardless of the machine count. As a result, throughput naturally increases as more cores and SSD bandwidth become available. 
For the \bigann{}, \deep{}, and \msspacev{} datasets, \name{} effectively exploits the data locality provided by graph partitioning, scaling better than both ScatterGather and DistributedANN as the cluster size increases to 10 servers.

However, as \name{} partitions shrink with an increasing number of servers, consecutive search steps are less likely to reside on the same node. This results in the increased inter-partition hop count observed in \autoref{fig:hops_vs_inter_hops}, preventing \name{} from achieving optimal linear scaling. Furthermore, for the \texttwoimage{} dataset, the performance of \name{} converges with DistributedANN for the reasons previously discussed in \autoref{subsec:multi_server_throughput}.

\subsection{Effects of Varying $K$}
\label{subsec:vary_k}
\begin{figure}[h]
    \centering
    \includegraphics[width=0.8\linewidth]{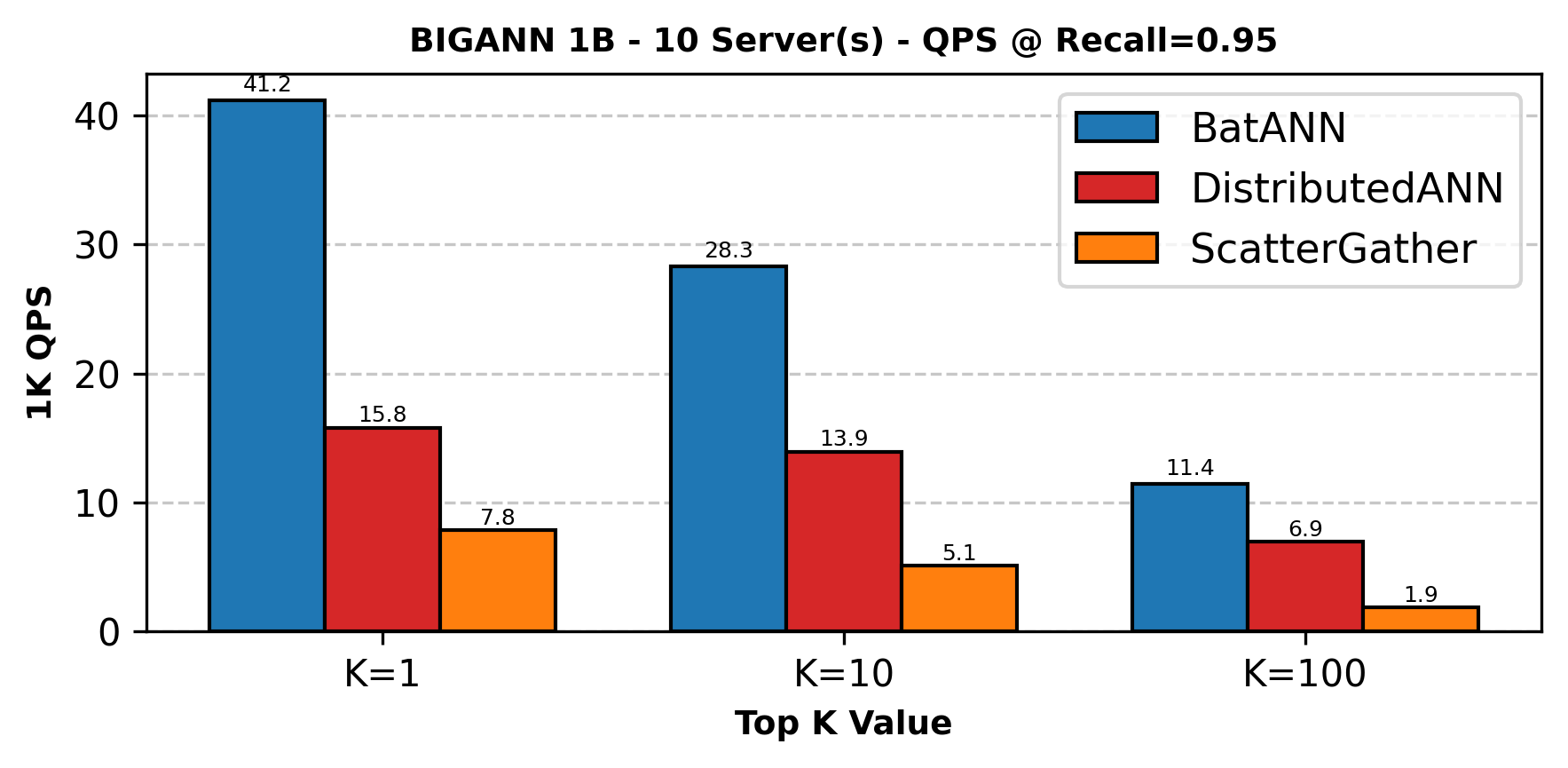}
\caption{QPS vs. recall on \bigann{} 1B 10 servers for $k=1,10,100$. \name{} achieves superior performance compared to all other methods at all $k$ values.}
\label{fig:qps_recall_bigann_1B_10_server_K_1_10_100_bar}
\end{figure}

To evaluate the impact of the varying $K$, we measured system throughput for $K = 1, 10, 100$ on the \bigann{} 1B dataset using the 10-server configuration. As shown in \autoref{fig:qps_recall_bigann_1B_10_server_K_1_10_100_bar}, the relative performance ranking among \name{}, ScatterGather, and DistributedANN remains consistent, with \name{} maintaining the highest throughput at a 0.95 recall target across all $K$ values. For the other datasets, \name{} continues to outperform both baselines across all $K$ values, although this performance gap narrows on the harder \texttwoimage{} dataset for 100M points at 0.7 recall@10, mirroring the trends discussed in \autoref{subsec:multi_server_throughput}.

The performance gap between \name{} and DistributedANN narrows as we increase $K$ because \name{} uses an approximation of beam search, with the help of \autoref{alg:distributed-beam}. DistributedANN faithfully executes beam search by having an orchestration server issue scoring queries in place of the direct disk I/O required by beam search (\autoref{alg:beam_search}). 
To get the approximate $K$ nearest neighbors, we re-rank the candidate set of size $L$ and take the best $K$ results. Because of the imprecision of the quantized distances, a larger candidate set will have more of the true $K$ nearest neighbors. This causes any given value of $L$ to become less effective for larger $K$, so the gap between \autoref{alg:distributed-beam} and beam search is more apparent.

\subsection{End-to-End Latency}
\label{subsec:latency}

We define end-to-end latency as the time between a client issuing a query and receiving the corresponding result. Prior single-server systems and evaluations \cite{jayaram_subramanya_diskann_2019, guoAchievingLowLatencyGraphBased2025} typically issue queries within the same search process using OMP dynamic scheduling and measure latency as the duration of the search function call. This methodology does not translate cleanly to a distributed setting where queues necessarily have to be used.

Accordingly, we split up our experiments into 2 sections. First we compare end-to-end latency for \name{}, DistributedANN, and ScatterGather at 1000 QPS to see how the systems perform with no queueing artifacts. We then look at how latency degrades (or doesn't) when we issue sends at rates close to max throughput.

\begin{figure}[h]
    \centering
    \includegraphics[width=0.65\linewidth]{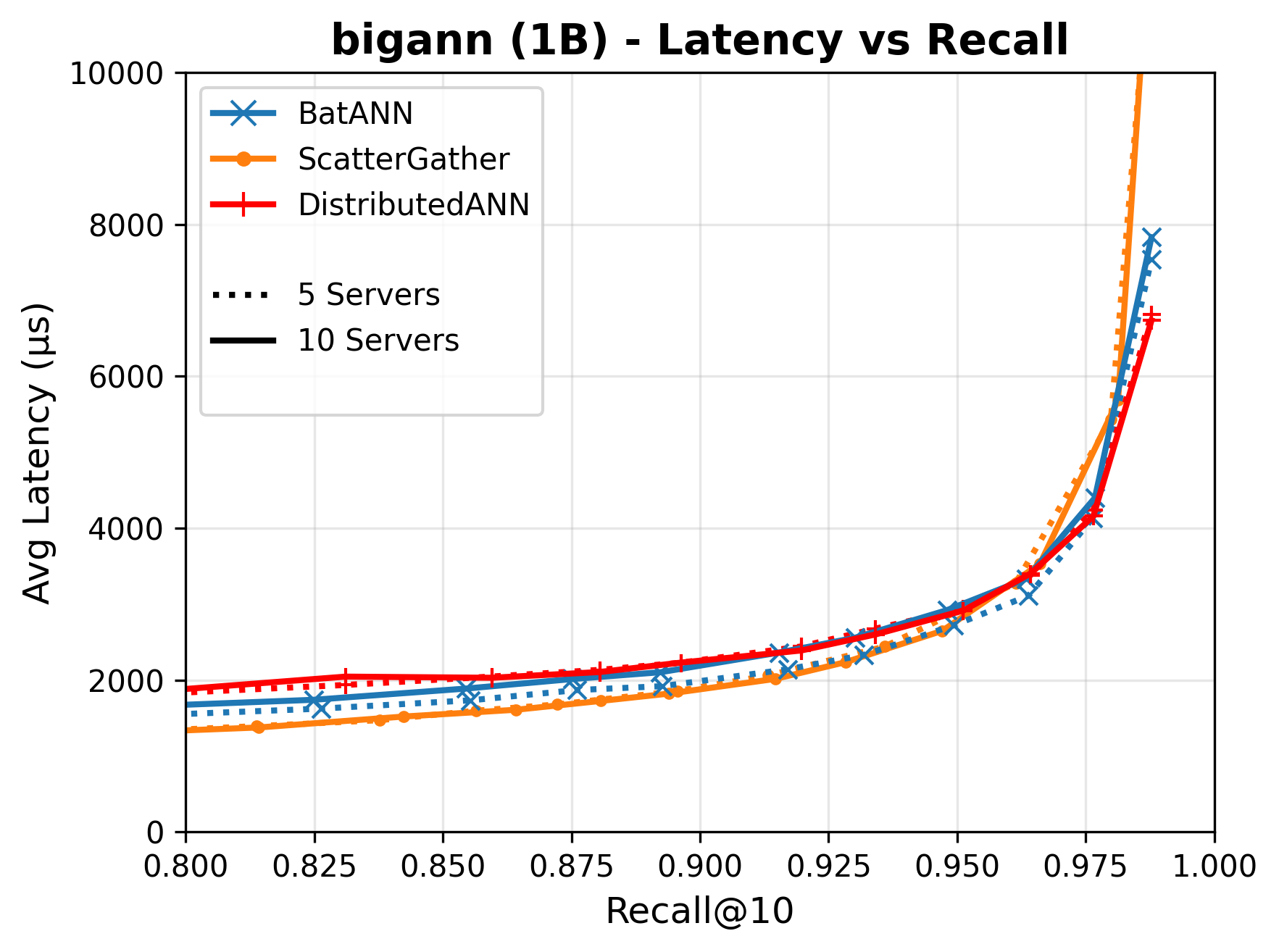}
    \caption{Latency Recall curve of \bigann{} 1B on 5, 10 server setup with at 1K send rate. We see a slight increase in \name{} latency when scaling up.}
    \label{fig:latency}
\end{figure}
\subsubsection{1K Send Rate}
As shown in \autoref{fig:latency}, \name{} achieves \textbf{end-to-end latency below 3 ms} at 0.95 recall@10 on both the 5-server and 10-server setups. Moreover, \name{}'s latency on 10 servers is only 8\% higher than on 5 servers.

On the 5-server setup, \name{} achieves the lowest mean latency of all methods at 2737 $\mu$s, compared to 2902 $\mu$s for ScatterGather and 2919.10 $\mu$s for DistributedANN. When scaling to 10 servers, the increased inter-server communication pushes \name{}'s mean latency to 2972.55 $\mu$s. At this scale, ScatterGather achieves the lowest mean latency (2781.57 $\mu$s), while DistributedANN remains relatively stable (2891.68 $\mu$s).



From \autoref{fig:dist_cmps_io_comparison}, we observe that the amount of disk I/O—the dominant bottleneck for latency in disk-based vector search \cite{guoAchievingLowLatencyGraphBased2025, jayaram_subramanya_diskann_2019}—remains nearly identical across both configurations. \autoref{fig:dist_cmps_io_comparison} further shows that the number of distance comparisons is almost identical in both settings. These similarities arise from the fact that both configurations search over the same global graph and therefore perform essentially the same amount of work per query. It also highlights the effectiveness of \autoref{alg:distributed-beam} at adapting the I/O pipeline in a distributed setting. The only meaningful difference is the higher number of inter-partition hops in the 10-server setup. Each hop requires transmitting a search state to another server, which incurs serialization and network-stack overhead. This additional communication cost accounts for the modest increase in end-to-end latency.

ScatterGather's latency is better than \name{} because it is only dependent on the time it takes to query the slowest partition. These disjoint partitions are smaller than the global index, which makes them slightly faster to query. For the same reason, ScatterGather latency is lower for the 10 server configuration, as the partitions are half as large.

DistributedANN's latency remains fairly constant across both server configurations. Because the total number of hops remains unchanged for a global graph, the workload is simply spread in parallel across a larger number of scoring servers without adding more round-trips.

\subsubsection{Latency vs. Send Rate for 0.95 recall}
\begin{figure}[h]
    \centering
    \includegraphics[width=0.65\linewidth]{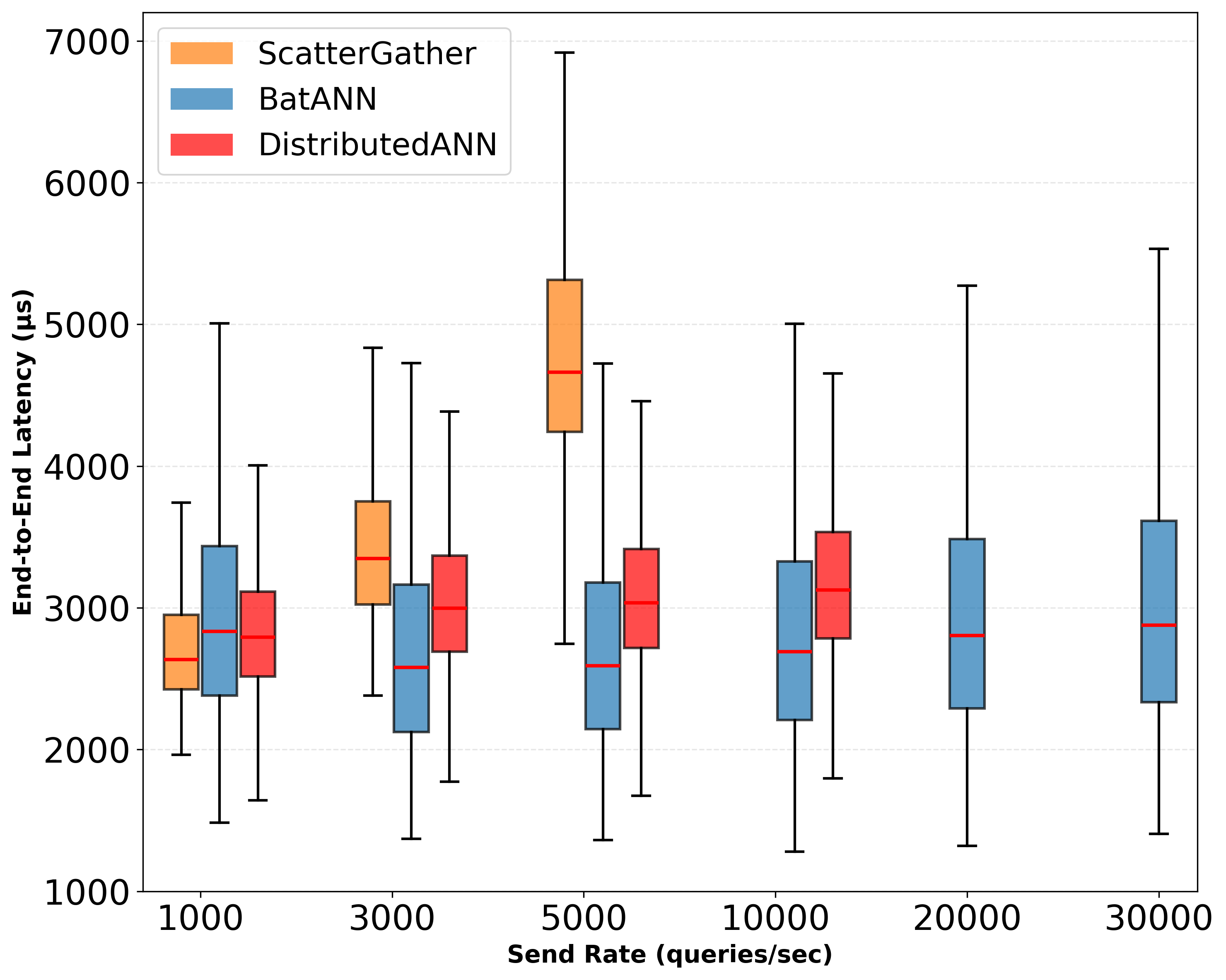}
    \caption{Box and whisker plots of end-to-end latency for  0.95 recall@10 on \bigann{} 1B on 10 servers.  \name{} is stable to 30000qps and has the lowest overall average latency.}
    \label{fig:latency_vs_send_rate}
\end{figure}
From \autoref{fig:qps_recall_1B}, we can see that throughput for ScatterGather and \name{} are around 5000 and 30000 QPS respectively. Hence, we choose to examine how latency changes as we increase the send rate up to these thresholds. As we can see in \autoref{fig:latency_vs_send_rate}, as send rate increases from 1000 to 5000, ScatterGather's mean and tail latency rise accordingly. We attribute this to the implicit synchronization overhead of ScatterGather. The client has to wait for results to arrive from all servers, so as we increase the send rate, if one server is struggling to keep up, then the latency is bottlenecked by this straggler. In contrast, \name{}'s mean latency remains fairly constant across all send rates and stays below 3ms. 

DistributedANN's mean latency is also fairly consistent at 3ms. However, it is worth noting that \name{} exhibits a slightly wider latency spread than DistributedANN. This variance is a consequence of \name{}'s  state-passing design; because different queries require a variable number of sequential inter-partition hops, they accumulate slightly different amounts of network and serialization overhead compared to the parallel, synchronized request-reply pattern utilized by DistributedANN.

\subsection{ScatterGather Top-$N$}
\label{subsec:scatter_gather_top_n}
\begin{figure}[h]
    \centering
    \includegraphics[width=\linewidth]{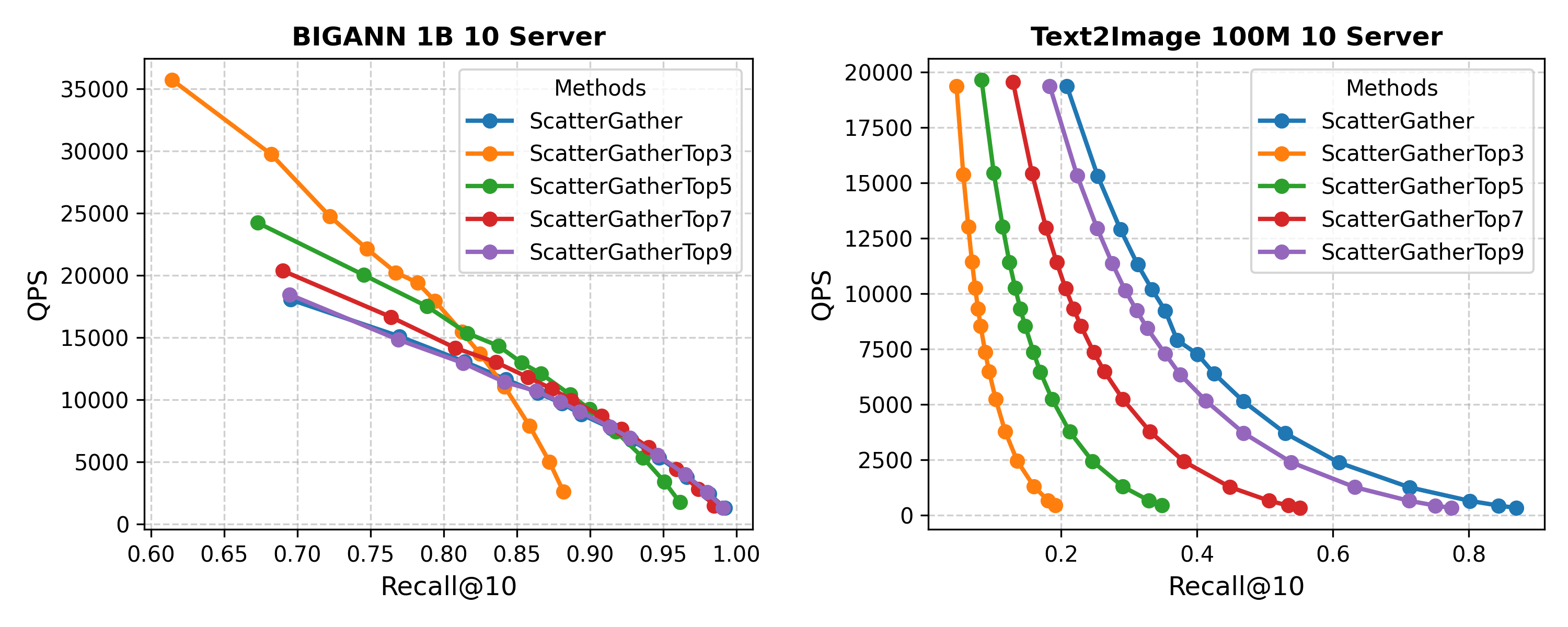}
    \caption{QPS vs recall@10 for \bigann{} 1B and \texttwoimage{} 100M for 10 servers with ScatterGather Top-$N$. Searching all partitions generally yielded highest throughput at higher recall values.}
    \label{fig:scatter_gather_top_n}
\end{figure}

If we partition the dataset using spatial partitioning techniques, then an additional approach to ScatterGather is to only send queries to the $N$ nearest partitions. For our implementation of ScatterGather Top-$N$, the client calculates the distance between the query and each partition centroid to select the $N$ closest partitions. Queries are then routed to each of these partitions.

From \autoref{fig:scatter_gather_top_n}, we see that for \bigann{} 1B, selecting the top $N$ partitions for scatter gather only improves throughput for lower recall values and not for recall@10 > 0.95. Many queries have true nearest neighbors split between several partitions, some of which may be excluded from the $N$ selected partitions. Additionally, for \texttwoimage{} 100M, which is a much harder dataset, the throughput of ScatterGather Top-N never matches searching all partitions, i.e., ScatterGather. At best, ScatterGather Top-$N$ throughput matches that of ScatterGather for higher recall values, and at worst, its throughput is not close to ScatterGather on any recall value, even as we search $N-1$ partitions.

\section{Related Work}
\label{sec:unrelated}

\subsection{Disk-based Vector Search}
To resolve DiskANN's poor spatial locality, Starling \cite{wang_starling_2024}, PageANN \cite{kangScalableDiskBasedApproximate2025}, and DiskANN++ \cite{niDiskANNEfficientPagebased2023} co-locate graph neighbors within the same disk sector. For high-dimensional datasets, Gorgeous \cite{yinGorgeousRevisitingData2025} packs pages with two-hop neighbor IDs to additional Disk I/O. Other works optimize the beam search procedure itself; PipeANN \cite{guoAchievingLowLatencyGraphBased2025} and AlayaLaser \cite{chenAlayaLaserEfficientIndex2026} dynamically adjust the I/O pipeline size during search and relax strict I/O Pipeline read ordering.

To reduce memory overhead, AiSAQ and AlayaLaser store quantized neighbor data directly on disk \cite{tatsuno_aisaq_2025,chenAlayaLaserEfficientIndex2026}. While this duplicates data and inflates overall storage requirements, AlayaLaser mitigates the footprint by exclusively storing SIMD-friendly principal component data. Furthermore, by introducing multiple beam-search optimizations, such as an adaptive I/O pipeline, AlayaLaser achieves throughput and latency superior to DiskANN and competitive with in-memory vector search libraries.

Cluster-based methods such as SPANN \cite{chenSPANNHighlyefficientBillionscale2021a} and BraveANN \cite{zhuBraveANNRobustApproximate2026} partition data into contiguous on-disk clusters. Search routes through an in-memory centroid index before linearly scanning the retrieved disk clusters. To maintain recall, boundary vectors are heavily replicated across clusters (8x).

Given the large existing  customer base on  traditional DBMS, there has been strong demand for integrating vector search into these systems. One of the most popular systems is pgvector which adds vector search to PostgreSQL. Shim et al. propose SSD-based optimizations to pgvector that achieve competitive throughput with DiskANN \cite{shimTurbochargingVectorDatabases2025}. Another paper, PostgreSQL-V, decouples the internal components of PostgreSQL from vector search and introduces lightweight consistency mechanisms to utilize external vector libraries without storing vector data in PostgreSQL \cite{liuFastVectorSearch2026}.

\subsection{Other Distributed Vector Search Systems}
\label{sec:prior-distributed}

Both purpose-built vector databases and traditional distributed databases commonly rely on a scatter--gather paradigm to enable distributed vector search. Systems like Milvus \cite{wang_milvus_2021, wang_starling_2024} and Pinecone \cite{ingber2025accurate} partition data into independent shards that are queried in parallel. Similarly, the NoSQL Azure Cosmos DB \cite{upretiCostEffectiveLowLatency2025} queries sharded DiskANN indices via scatter--gather, and TigerVector \cite{liuTigerVectorSupportingVector2025a} embeds distributed scatter--gather vector search natively into the TigerGraph graph database.  

SPIRE~\cite{xuScalableDistributedVector2025a} adapts a cluster-based method (SPANN) for distributed settings using recursive bottom-up hierarchical clustering of data with balanced k-means until the topmost layer fits entirely in memory. The technique enables much better throughput than ScatterGather with provably bounded latency. While SPIRE is very promising, its code base was not publically available, so we were unable to compare against it.

As datacenters are shifting to disaggregated architectures to enable better resource utilization and scaling, there have been recent developments in the space of disaggregated memory vector search systems. Cheng et al. explored tradeoffs between CXL, RDMA, and NVMe as second tier memory for cluster and graph based approaches \cite{chengCharacterizingDilemmaPerformance2024}. SHINE and d-HNSW explore RDMA based disaggregated memory HNSW graph systems \cite{widmoserSHINEScalableHNSW2025,liuEfficientVectorSearch2025}.

\section{Discussion and Future Work}
Our work leaves open a number of topics for future investigation:

\textbf{Reducing Message Size:} 
Currently, each state transfer includes both the beam and the full result set, which can reach 3.2 KB for large $L$ (200–400) needed for high recall. Because this data is only used for the final reranking, servers could instead send partial results directly to the client during state transfers. The client would then aggregate and rerank them, similar to ScatterGather. However, our implementation of this approach yielded lower recall than the default \name{} system.

\textbf{Exploration of Disk-Based Optimizations:} 
To overcome the in-memory limitation of storing all quantized vector data in each server, we can rely on the AlayaLaser as discussed in \autoref{sec:unrelated} to store this data on disk instead. AlayaLaser's throughput and latency are superior to DiskANN especially on high-dimensional datasets and even resembles that of in-memory vector search libraries. Using their disk layout and adaptive I/O pipeline solution would solve the in-memory bottleneck of our system and improve its throughput and latency as well. 

\textbf{Distributed Global Graph Updates:} 
A wealth of prior work has studied ways to support in-place updates to search graphs without compromising the quality of the index or periodically re-indexing \cite{xuInPlaceUpdatesGraph2025, xu2023spfresh, singh_freshdiskann_2021, guo_odinann_nodate}. The distributed nature of our system presents unique challenges in this area. After removing a point from the graph, in-neighbors of that point which may be spread across several nodes need to update their neighborhoods as not to route search to a point which no longer exists.

\textbf{Efficient Fault Tolerance:}
Our current system is not robust to node failures.  Our plan is to port \name{} to run on the sharded storage framework provided by Derecho~\cite{derecho_2019}, which offers efficient fault-tolerant replication over RDMA.

\textbf{Network Accelerators:} 
Previous work on global distributed graph indices has leveraged RDMA and CXL \cite{zhi_towards_2025, jangCXLANNSSoftwareHardwareCollaborative2023}, which require more specialized and expensive hardware than our TCP-based approach. Because our design is based on a high-speed asynchronous data-relaying scheme centered on query states that are only a few kilobytes in size (an object size at which datacenter TCP performs well), it is unclear that RDMA would offer a significant speedup. Extensions to this work that would benefit more from the performance characteristics of RDMA are a promising future direction.


\section{Conclusion}

We present \name{}, a novel system for distributed disk-based ANNS. The core innovation of \name{} is that we query a global graph by sending query state between machines when beam search expands a node on another server. This allows us to achieve better latency, faster communication, and improved locality, while preserving the number of disk reads and distance comparisons compared to a single node baseline.





\begin{acks}
 This work was supported by funds provided by Microsoft and IBM. We thank CloudLab and their supporters for access to machines. We also thank Alicia Yang for valuable input and technical support.
\end{acks}


\bibliographystyle{acm}
\bibliography{references}

@article{yu_a_malkov_efficient_2020,
    title = {Efficient and {Robust} {Approximate} {Nearest} {Neighbor} {Search} {Using} {Hierarchical} {Navigable} {Small} {World} {Graphs}},
    volume = {42},
    doi = {10.1109/tpami.2018.2889473},
    number = {4},
    journal = {IEEE Transactions on Pattern Analysis and Machine Intelligence},
    author = {{Yu. A. Malkov} and {Yu. A. Malkov} and Malkov, Yu. A. and {D. A. Yashunin} and Yashunin, D. A.},
    month = apr,
    year = {2020},
    doi = {10.1109/tpami.2018.2889473},
    pmid = {30602420},
    pages = {824--836},
}

@inproceedings{lewis2020rag,
    title = {Retrieval-augmented generation for knowledge-intensive {NLP} tasks},
    volume = {33},
    url = {https://proceedings.neurips.cc/paper_files/paper/2020/file/6b493230205f780e1bc26945df7481e5-Paper.pdf},
    booktitle = {Advances in neural information processing systems},
    publisher = {Curran Associates, Inc.},
    author = {Lewis, Patrick and Perez, Ethan and Piktus, Aleksandra and Petroni, Fabio and Karpukhin, Vladimir and Goyal, Naman and Küttler, Heinrich and Lewis, Mike and Yih, Wen-tau and Rocktäschel, Tim and Riedel, Sebastian and Kiela, Douwe},
    editor = {Larochelle, H. and Ranzato, M. and Hadsell, R. and Balcan, M.F. and Lin, H.},
    year = {2020},
    pages = {9459--9474},
}

@inproceedings{adams_distributedann_2025,
    title = {{DistributedANN}: {Efficient} {Scaling} of a {Single} {DiskANN} {Graph} {Across} {Thousands} of {Computers}},
    shorttitle = {{DistributedANN}},
    url = {https://openreview.net/forum?id=6AEsfCLRm3},
    language = {en},
    urldate = {2025-07-16},
    author = {Adams, Philip and Li, Menghao and Zhang, Shi and Tan, Li and Chen, Qi and Li, Mingqin and Li, Zengzhong and Risvik, Knut Magne and Simhadri, Harsha Vardhan},
    month = jul,
    booktitle = {The 1st Workshop on Vector Databases},
    year = {2025},
}

@inproceedings{huang_learning_2013,
    address = {San Francisco California USA},
    title = {Learning deep structured semantic models for web search using clickthrough data},
    isbn = {978-1-4503-2263-8},
    url = {https://dl.acm.org/doi/10.1145/2505515.2505665},
    doi = {10.1145/2505515.2505665},
    language = {en},
    urldate = {2025-11-06},
    booktitle = {Proceedings of the 22nd {ACM} international conference on {Information} \& {Knowledge} {Management}},
    publisher = {ACM},
    author = {Huang, Po-Sen and He, Xiaodong and Gao, Jianfeng and Deng, Li and Acero, Alex and Heck, Larry},
    month = oct,
    year = {2013},
    pages = {2333--2338},
}

@article{derecho_2019,
author = {Jha, Sagar and Behrens, Jonathan and Gkountouvas, Theo and Milano, Mae and Song, Weijia and Tremel, Edward and Renesse, Robbert Van and Zink, Sydney and Birman, Kenneth P.},
title = {Derecho: Fast State Machine Replication for Cloud Services},
year = {2019},
issue_date = {May 2018},
publisher = {Association for Computing Machinery},
address = {New York, NY, USA},
volume = {36},
number = {2},
issn = {0734-2071},
url = {https://doi.org/10.1145/3302258},
doi = {10.1145/3302258},
journal = {ACM Trans. Comput. Syst.},
month = apr,
articleno = {4},
numpages = {49}
}

@inproceedings{radford_learning_2021,
    title = {Learning {Transferable} {Visual} {Models} {From} {Natural} {Language} {Supervision}},
    url = {https://proceedings.mlr.press/v139/radford21a.html},
    language = {en},
    urldate = {2025-11-06},
    booktitle = {Proceedings of the 38th {International} {Conference} on {Machine} {Learning}},
    publisher = {PMLR},
    author = {Radford, Alec and Kim, Jong Wook and Hallacy, Chris and Ramesh, Aditya and Goh, Gabriel and Agarwal, Sandhini and Sastry, Girish and Askell, Amanda and Mishkin, Pamela and Clark, Jack and Krueger, Gretchen and Sutskever, Ilya},
    month = jul,
    year = {2021},
    note = {ISSN: 2640-3498},
    pages = {8748--8763},
}

@inproceedings{jayaram_subramanya_diskann_2019,
    title = {{DiskANN}: {Fast} {Accurate} {Billion}-point {Nearest} {Neighbor} {Search} on a {Single} {Node}},
    volume = {32},
    shorttitle = {{DiskANN}},
    url = {https://papers.nips.cc/paper_files/paper/2019/hash/09853c7fb1d3f8ee67a61b6bf4a7f8e6-Abstract.html},
    urldate = {2024-11-15},
    booktitle = {Advances in {Neural} {Information} {Processing} {Systems}},
    publisher = {Curran Associates, Inc.},
    author = {Jayaram Subramanya, Suhas and Devvrit, Fnu and Simhadri, Harsha Vardhan and Krishnawamy, Ravishankar and Kadekodi, Rohan},
    year = {2019},
}

@article{wang_starling_2024,
    title = {Starling: {An} {I}/{O}-{Efficient} {Disk}-{Resident} {Graph} {Index} {Framework} for {High}-{Dimensional} {Vector} {Similarity} {Search} on {Data} {Segment}},
    volume = {2},
    issn = {2836-6573},
    shorttitle = {Starling},
    url = {http://arxiv.org/abs/2401.02116},
    doi = {10.1145/3639269},
    number = {1},
    urldate = {2025-06-06},
    journal = {Proceedings of the ACM on Management of Data},
    author = {Wang, Mengzhao and Xu, Weizhi and Yi, Xiaomeng and Wu, Songlin and Peng, Zhangyang and Ke, Xiangyu and Gao, Yunjun and Xu, Xiaoliang and Guo, Rentong and Xie, Charles},
    month = mar,
    year = {2024},
    note = {arXiv:2401.02116 [cs]},
    pages = {1--27},
}

@inproceedings{arya_approximate_1993,
    address = {Austin, Texas, USA},
    series = {Soda '93},
    title = {Approximate nearest neighbor queries in fixed dimensions},
    isbn = {0-89871-313-7},
    booktitle = {Proceedings of the fourth annual {ACM}-{SIAM} symposium on discrete algorithms},
    publisher = {Society for Industrial and Applied Mathematics},
    author = {Arya, Sunil and Mount, David M.},
    year = {1993},
    note = {Number of pages: 10
tex.address: USA},
    pages = {271--280},
}

@article{jegou_product_2011,
    title = {Product {Quantization} for {Nearest} {Neighbor} {Search}},
    volume = {33},
    copyright = {https://ieeexplore.ieee.org/Xplorehelp/downloads/license-information/IEEE.html},
    issn = {0162-8828},
    url = {http://ieeexplore.ieee.org/document/5432202/},
    doi = {10.1109/TPAMI.2010.57},
    number = {1},
    urldate = {2024-11-14},
    journal = {IEEE Transactions on Pattern Analysis and Machine Intelligence},
    author = {Jégou, H and Douze, M and Schmid, C},
    month = jan,
    year = {2011},
    pages = {117--128},
}

@misc{tatsuno_aisaq_2025,
    title = {{AiSAQ}: {All}-in-{Storage} {ANNS} with {Product} {Quantization} for {DRAM}-free {Information} {Retrieval}},
    shorttitle = {{AiSAQ}},
    url = {http://arxiv.org/abs/2404.06004},
    doi = {10.48550/arXiv.2404.06004},
    urldate = {2025-07-16},
    publisher = {arXiv},
    author = {Tatsuno, Kento and Miyashita, Daisuke and Ikeda, Taiga and Ishiyama, Kiyoshi and Sumiyoshi, Kazunari and Deguchi, Jun},
    month = feb,
    year = {2025},
    note = {arXiv:2404.06004 [cs]},
}

@inproceedings{wang_milvus_2021,
    address = {Virtual Event China},
    title = {Milvus: {A} {Purpose}-{Built} {Vector} {Data} {Management} {System}},
    isbn = {978-1-4503-8343-1},
    shorttitle = {Milvus},
    url = {https://dl.acm.org/doi/10.1145/3448016.3457550},
    doi = {10.1145/3448016.3457550},
    language = {en},
    urldate = {2024-11-14},
    booktitle = {Proceedings of the 2021 {International} {Conference} on {Management} of {Data}},
    publisher = {ACM},
    author = {Wang, Jianguo and Yi, Xiaomeng and Guo, Rentong and Jin, Hai and Xu, Peng and Li, Shengjun and Wang, Xiangyu and Guo, Xiangzhou and Li, Chengming and Xu, Xiaohai and Yu, Kun and Yuan, Yuxing and Zou, Yinghao and Long, Jiquan and Cai, Yudong and Li, Zhenxiang and Zhang, Zhifeng and Mo, Yihua and Gu, Jun and Jiang, Ruiyi and Wei, Yi and Xie, Charles},
    month = jun,
    year = {2021},
    pages = {2614--2627},
}

@inproceedings{ingber2025accurate,
    title = {Accurate and efficient metadata filtering in pinecone’s serverless vector database},
    url = {https://openreview.net/forum?id=UXq4z6GGYP},
    booktitle = {The 1st workshop on vector databases},
    author = {Ingber, Amir and Liberty, Edo},
    year = {2025},
}

@misc{dilocker_weaviate_2025,
    title = {Weaviate},
    copyright = {BSD-3-Clause},
    url = {https://github.com/weaviate/weaviate},
    urldate = {2025-11-04},
    author = {Dilocker, Etienne and van Luijt, Bob and Voorbach, Byron and Hasan, Mohd Shukri and Rodriguez, Abdel and Kulawiak, Dirk Alexander and Antas, Marcin and Duckworth, Parker},
    month = nov,
    year = {2025},
    note = {original-date: 2016-03-30T15:03:17Z},
}

@misc{deng_pyramid_2019,
    title = {Pyramid: {A} {General} {Framework} for {Distributed} {Similarity} {Search}},
    shorttitle = {Pyramid},
    url = {http://arxiv.org/abs/1906.10602},
    language = {en},
    urldate = {2024-11-14},
    publisher = {arXiv},
    author = {Deng, Shiyuan and Yan, Xiao and Ng, Kelvin K. W. and Jiang, Chenyu and Cheng, James},
    month = jun,
    year = {2019},
    note = {arXiv:1906.10602 [cs]},
}

@misc{zhi_towards_2025,
    title = {Towards {Efficient} and {Scalable} {Distributed} {Vector} {Search} with {RDMA}},
    url = {http://arxiv.org/abs/2507.06653},
    doi = {10.48550/arXiv.2507.06653},
    urldate = {2025-07-16},
    publisher = {arXiv},
    author = {Zhi, Xiangyu and Chen, Meng and Yan, Xiao and Lu, Baotong and Li, Hui and Zhang, Qianxi and Chen, Qi and Cheng, James},
    month = jul,
    year = {2025},
    note = {arXiv:2507.06653 [cs]},
}

@inproceedings{jangCXLANNSSoftwareHardwareCollaborative2023,
  title = {{{CXL-ANNS}}: {{Software-Hardware}} Collaborative Memory Disaggregation and Computation for {{Billion-Scale}} Approximate Nearest Neighbor Search},
  booktitle = {2023 {{USENIX}} Annual Technical Conference ({{USENIX ATC}} 23)},
  author = {Jang, Junhyeok and Choi, Hanjin and Bae, Hanyeoreum and Lee, Seungjun and Kwon, Miryeong and Jung, Myoungsoo},
  date = {2023-07},
  year = {2023},
  pages = {585--600},
  publisher = {USENIX Association},
  location = {Boston, MA},
  url = {https://www.usenix.org/conference/atc23/presentation/jang},
  isbn = {978-1-939133-35-9}
}

@misc{gottesburen_unleashing_2024,
    title = {Unleashing {Graph} {Partitioning} for {Large}-{Scale} {Nearest} {Neighbor} {Search}},
    url = {http://arxiv.org/abs/2403.01797},
    doi = {10.48550/arXiv.2403.01797},
    urldate = {2024-11-24},
    publisher = {arXiv},
    author = {Gottesbüren, Lars and Dhulipala, Laxman and Jayaram, Rajesh and Lacki, Jakub},
    month = mar,
    year = {2024},
    note = {arXiv:2403.01797},
}

@inproceedings{zhang_vbase_2023,
    author = {Qianxi Zhang and Shuotao Xu and Qi Chen and Guoxin Sui and Jiadong Xie and Zhizhen Cai and Yaoqi Chen and Yinxuan He and Yuqing Yang and Fan Yang and Mao Yang and Lidong Zhou},
    title = {{VBASE}: Unifying Online Vector Similarity Search and Relational Queries via Relaxed Monotonicity},
    booktitle = {17th USENIX Symposium on Operating Systems Design and Implementation (OSDI 23)},
    year = {2023},
    isbn = {978-1-939133-34-2},
    address = {Boston, MA},
    pages = {377--395},
    url = {https://www.usenix.org/conference/osdi23/presentation/zhang-qianxi},
    publisher = {USENIX Association},
    month = jul
}

@article{xuTwostageRoutingOptimized2021,
  title = {Two-Stage Routing with Optimized Guided Search and Greedy Algorithm on Proximity Graph},
  author = {Xu, Xiaoliang and Wang, Mengzhao and Wang, Yuxiang and Ma, Dingcheng},
  date = {2021-10-11},
  year = {2021},
  journal = {Knowledge-Based Systems},
  shortjournal = {Knowledge-Based Systems},
  volume = {229},
  pages = {107305},
  issn = {0950-7051},
  doi = {10.1016/j.knosys.2021.107305},
  url = {https://www.sciencedirect.com/science/article/pii/S0950705121005670},
  urldate = {2025-05-15}
}

@misc{zeromq_lib,
    title = {{ZeroMQ}: {The} intelligent transport layer},
    url = {https://zeromq.org/},
    author = {{The ZeroMQ Community}},
    year = {2025},
}

@article{desrochers_fast_2014,
    title = {A {Fast} {General} {Purpose} {Lock}-{Free} {Queue} for {C}++},
    url = {https://moodycamel.com/blog/2014/a-fast-general-purpose-lock-free-queue-for-c++},
    journal = {moodycamel.com},
    author = {Desrochers, Cameron},
    year = {2014},
}

@inproceedings{duplyakin_design_2019,
    title = {The design and operation of {CloudLab}},
    url = {https://www.flux.utah.edu/paper/duplyakin-atc19},
    booktitle = {Proceedings of the {USENIX} annual technical conference ({ATC})},
    author = {Duplyakin, Dmitry and Ricci, Robert and Maricq, Aleksander and Wong, Gary and Duerig, Jonathon and Eide, Eric and Stoller, Leigh and Hibler, Mike and Johnson, David and Webb, Kirk and Akella, Aditya and Wang, Kuangching and Ricart, Glenn and Landweber, Larry and Elliott, Chip and Zink, Michael and Cecchet, Emmanuel},
    month = jul,
    year = {2019},
    pages = {1--14},
}

@misc{simhadri2024resultsbigannneurips23,
    title = {Results of the big {ANN}: {NeurIPS}'23 competition},
    url = {https://arxiv.org/abs/2409.17424},
    author = {Simhadri, Harsha Vardhan and Aumüller, Martin and Ingber, Amir and Douze, Matthijs and Williams, George and Manohar, Magdalen Dobson and Baranchuk, Dmitry and Liberty, Edo and Liu, Frank and Landrum, Ben and Karjikar, Mazin and Dhulipala, Laxman and Chen, Meng and Chen, Yue and Ma, Rui and Zhang, Kai and Cai, Yuzheng and Shi, Jiayang and Chen, Yizhuo and Zheng, Weiguo and Wan, Zihao and Yin, Jie and Huang, Ben},
    year = {2024},
    note = {arXiv: 2409.17424 [cs.IR]},
}

@inproceedings{manoharParlayANNScalableDeterministic2024,
  title = {{{ParlayANN}}: {{Scalable}} and Deterministic Parallel Graph-Based Approximate Nearest Neighbor Search Algorithms},
  booktitle = {Proceedings of the 29th {{ACM SIGPLAN}} Annual Symposium on Principles and Practice of Parallel Programming},
  author = {Manohar, Magdalen Dobson and Shen, Zheqi and Blelloch, Guy and Dhulipala, Laxman and Gu, Yan and Simhadri, Harsha Vardhan and Sun, Yihan},
  year = {2024},
  series = {{{PPoPP}} '24},
  pages = {270--285},
  publisher = {Association for Computing Machinery},
  address = {New York, NY, USA},
  location = {Edinburgh, United Kingdom},
  doi = {10.1145/3627535.3638475},
  url = {https://doi.org/10.1145/3627535.3638475},
  isbn = {979-8-4007-0435-2},
  pagetotal = {16}
}

@article{aumuller_ann-benchmarks_2020,
    title = {{ANN}-{Benchmarks}: {A} benchmarking tool for approximate nearest neighbor algorithms},
    volume = {87},
    issn = {0306-4379},
    shorttitle = {{ANN}-{Benchmarks}},
    url = {https://www.sciencedirect.com/science/article/pii/S0306437918303685},
    doi = {10.1016/j.is.2019.02.006},
    urldate = {2025-05-27},
    journal = {Information Systems},
    author = {Aumüller, Martin and Bernhardsson, Erik and Faithfull, Alexander},
    month = jan,
    year = {2020},
    pages = {101374},
}

@inproceedings{guoAchievingLowLatencyGraphBased2025,
  title = {Achieving \{{{Low-Latency}}\} \{{{Graph-Based}}\} {{Vector Search}} via {{Aligning}} \{{{Best-First}}\} {{Search Algorithm}} with \{{{SSD}}\}},
  author = {Guo, Hao and Lu, Youyou},
  year = {2025},
  pages = {171--186},
  url = {https://www.usenix.org/conference/osdi25/presentation/guo},
  urldate = {2025-07-07},
  booktitle = {19th {{USENIX Symposium}} on {{Operating Systems Design}} and {{Implementation}} ({{OSDI}} 25)},
  isbn = {978-1-939133-47-2},
  langid = {english}
}

@article{xuInplaceUpdatesGraph2025,
  title = {In-Place Updates of a Graph Index for Streaming Approximate Nearest Neighbor Search},
  author = {Xu, Haike and Manohar, Magdalen Dobson and Bernstein, Philip A. and Chandramouli, Badrish and Wen, Richard and Simhadri, Harsha Vardhan},
  year = 2025,
  month = feb,
  journal = {CoRR},
  volume = {abs/2502.13826},
  cdate = {1738368000000},
  publtype = {informal}
}

@inproceedings{xu2023spfresh,
    title = {{SPFresh}: {Incremental} in-place update for billion-scale vector search},
    booktitle = {Proceedings of the 29th symposium on operating systems principles},
    author = {Xu, Yuming and Liang, Hengyu and Li, Jin and Xu, Shuotao and Chen, Qi and Zhang, Qianxi and Li, Cheng and Yang, Ziyue and Yang, Fan and Yang, Yuqing and {others}},
    year = {2023},
    pages = {545--561},
}

@misc{singh_freshdiskann_2021,
    title = {{FreshDiskANN}: {A} {Fast} and {Accurate} {Graph}-{Based} {ANN} {Index} for {Streaming} {Similarity} {Search}},
    shorttitle = {{FreshDiskANN}},
    url = {http://arxiv.org/abs/2105.09613},
    language = {en},
    urldate = {2024-11-14},
    publisher = {arXiv},
    author = {Singh, Aditi and Subramanya, Suhas Jayaram and Krishnaswamy, Ravishankar and Simhadri, Harsha Vardhan},
    month = may,
    year = {2021},
    note = {arXiv:2105.09613 [cs]},
}

@article{guo_odinann_nodate,
    title = {{OdinANN}: {Direct} {Insert} for {Consistently} {Stable} {Performance} in {Billion}-{Scale} {Graph}-{Based} {Vector} {Search}},
    journal={24th USENIX Conference on File and Storage Technologies (FAST '26)},
    year={2026},
    language = {en},
    author = {Guo, Hao and Lu, Youyou},
}

@article{chavez_searching_2001,
    title = {Searching in metric spaces},
    volume = {33},
    issn = {0360-0300},
    url = {https://doi.org/10.1145/502807.502808},
    doi = {10.1145/502807.502808},
    number = {3},
    journal = {Acm Computing Surveys},
    author = {Ch{\'a}vez, Edgar and Navarro, Gonzalo and Baeza-Yates, Ricardo and Marroqu{\'{\i}}n, Jos{\'e} Luis},
    month = sep,
    year = {2001},
    pages = {273--321},
}

@inproceedings{dongEfficientKnearestNeighbor2011,
  title = {Efficient K-Nearest Neighbor Graph Construction for Generic Similarity Measures},
  booktitle = {Proceedings of the 20th International Conference on {{World}} Wide Web},
  author = {Dong, Wei and Moses, Charikar and Li, Kai},
  date = {2011-03-28},
  year = {2011},
  pages = {577--586},
  publisher = {ACM},
  location = {Hyderabad India},
  doi = {10.1145/1963405.1963487},
  url = {https://dl.acm.org/doi/10.1145/1963405.1963487},
  urldate = {2025-12-01},
  eventtitle = {{{WWW}} '11: 20th {{International World Wide Web Conference}}},
  isbn = {978-1-4503-0632-4},
  langid = {english},
}

@article{andreQuickerADCUnlocking2021,
  title = {Quicker {{ADC}} : {{Unlocking}} the Hidden Potential of {{Product Quantization}} with {{SIMD}}},
  shorttitle = {Quicker {{ADC}}},
  author = {Andr{\'e}, Fabien and Kermarrec, Anne-Marie and Scouarnec, Nicolas Le},
  year = 2021,
  month = may,
  journal = {IEEE Transactions on Pattern Analysis and Machine Intelligence},
  volume = {43},
  number = {5},
  eprint = {1812.09162},
  primaryclass = {cs},
  pages = {1666--1677},
  issn = {0162-8828, 2160-9292, 1939-3539},
  doi = {10.1109/TPAMI.2019.2952606},
  urldate = {2025-05-15},
  archiveprefix = {arXiv}
}

@article{johnsonBillionScaleSimilaritySearch2021,
  title = {Billion-{{Scale Similarity Search}} with {{GPUs}}},
  author = {Johnson, Jeff and Douze, Matthijs and J{\'e}gou, Herv{\'e}},
  year = 2021,
  month = jul,
  journal = {IEEE Transactions on Big Data},
  volume = {7},
  number = {3},
  pages = {535--547},
  issn = {2332-7790},
  doi = {10.1109/TBDATA.2019.2921572},
  urldate = {2025-12-01}
}

@misc{kangScalableDiskBasedApproximate2025,
  title = {Scalable {{Disk-Based Approximate Nearest Neighbor Search}} with {{Page-Aligned Graph}}},
  author = {Kang, Dingyi and Jiang, Dongming and Yang, Hanshen and Liu, Hang and Li, Bingzhe},
  year = 2025,
  month = sep,
  number = {arXiv:2509.25487},
  eprint = {2509.25487},
  primaryclass = {cs},
  publisher = {arXiv},
  doi = {10.48550/arXiv.2509.25487},
  urldate = {2025-10-01},
  archiveprefix = {arXiv}
}

@article{ahaltCompetitiveLearningAlgorithms1990,
  title = {Competitive Learning Algorithms for Vector Quantization},
  author = {Ahalt, Stanley C. and Krishnamurthy, Ashok K. and Chen, Prakoon and Melton, Douglas E.},
  year = 1990,
  month = jan,
  journal = {Neural Networks},
  volume = {3},
  number = {3},
  pages = {277--290},
  issn = {0893-6080},
  doi = {10.1016/0893-6080(90)90071-R},
  urldate = {2025-12-01},
}

@article{upretiCostEffectiveLowLatency2025,
  title = {Cost-{{Effective}}, {{Low Latency Vector Search}} with {{Azure Cosmos DB}}},
  author = {Upreti, Nitish and Simhadri, Harsha Vardhan and Sundar, Hari Sudan and Sundaram, Krishnan and Boshra, Samer and Perumalswamy, Balachandar and Atri, Shivam and Chisholm, Martin and Singh, Revti Raman and Yang, Greg and Hass, Tamara and Dudhey, Nitesh and Pattipaka, Subramanyam and Hildebrand, Mark and Manohar, Magdalen and Moffitt, Jack and Xu, Haiyang and Datha, Naren and Gupta, Suryansh and Krishnaswamy, Ravishankar and Gupta, Prashant and Sahu, Abhishek and Varada, Hemeswari and Barthwal, Sudhanshu and Mor, Ritika and Codella, James and Cooper, Shaun and Pilch, Kevin and Moreno, Simon and Kataria, Aayush and Kulkarni, Santosh and Deshpande, Neil and Sagare, Amar and Billa, Dinesh and Fu, Zishan and Vishal, Vipul},
  year = 2025,
  month = aug,
  journal = {Proc. VLDB Endow.},
  volume = {18},
  number = {12},
  pages = {5166--5183},
  issn = {2150-8097},
  doi = {10.14778/3750601.3750635},
  urldate = {2025-12-01},
}

@online{chengCharacterizingDilemmaPerformance2024,
  title = {Characterizing the {{Dilemma}} of {{Performance}} and {{Index Size}} in {{Billion-Scale Vector Search}} and {{Breaking It}} with {{Second-Tier Memory}}},
  author = {Cheng, Rongxin and Peng, Yifan and Wei, Xingda and Xie, Hongrui and Chen, Rong and Shen, Sijie and Chen, Haibo},
  date = {2024-05-07},
  eprint = {2405.03267},
  eprinttype = {arXiv},
  eprintclass = {cs},
  doi = {10.48550/arXiv.2405.03267},
  url = {http://arxiv.org/abs/2405.03267},
  urldate = {2025-06-04},
  pubstate = {prepublished},
  version = {2}
}

@inproceedings{shiScalableOverloadAwareGraphBased2025a,
  title = {Scalable {{Overload-Aware Graph-Based Index Construction}} for 10-{{Billion-Scale Vector Similarity Search}}},
  booktitle = {Companion {{Proceedings}} of the {{ACM}} on {{Web Conference}} 2025},
  author = {Shi, Yang and Sun, Yiping and Du, Jiaolong and Zhong, Xiaocheng and Wang, Zhiyong and Hu, Yao},
  date = {2025-05-23},
  series = {{{WWW}} '25},
  pages = {1303--1307},
  publisher = {Association for Computing Machinery},
  location = {New York, NY, USA},
  doi = {10.1145/3701716.3715576},
  url = {https://dl.acm.org/doi/10.1145/3701716.3715576},
  urldate = {2026-03-29},
  isbn = {979-8-4007-1331-6}
}

@misc{chenAlayaLaserEfficientIndex2026,
  title = {{{AlayaLaser}}: {{Efficient Index Layout}} and {{Search Strategy}} for {{Large-scale High-dimensional Vector Similarity Search}}},
  shorttitle = {{{AlayaLaser}}},
  author = {Chen, Weijian and Liu, Haotian and Deng, Yangshen and Xiang, Long and Huang, Liang and Li, Gezi and Tang, Bo},
  year = 2026,
  month = feb,
  number = {arXiv:2602.23342},
  eprint = {2602.23342},
  primaryclass = {cs},
  publisher = {arXiv},
  doi = {10.48550/arXiv.2602.23342},
  urldate = {2026-03-04},
  archiveprefix = {arXiv}
}

@article{shimTurbochargingVectorDatabases2025,
  title = {Turbocharging {{Vector Databases Using Modern SSDs}}},
  author = {Shim, Joobo and Oh, Jaewon and Roh, Hongchan and Do, Jaeyoung and Lee, Sang-Won},
  year = 2025,
  month = jul,
  journal = {Proceedings of the VLDB Endowment},
  volume = {18},
  number = {11},
  pages = {4710--4722},
  issn = {2150-8097},
  doi = {10.14778/3749646.3749724},
  urldate = {2026-03-30},
  langid = {english}
}

@misc{yinGorgeousRevisitingData2025,
  title = {Gorgeous: {{Revisiting}} the {{Data Layout}} for {{Disk-Resident High-Dimensional Vector Search}}},
  shorttitle = {Gorgeous},
  author = {Yin, Peiqi and Yan, Xiao and Zhou, Qihui and Li, Hui and Li, Xiaolu and Zhang, Lin and Wang, Meiling and Yao, Xin and Cheng, James},
  year = 2025,
  month = aug,
  number = {arXiv:2508.15290},
  eprint = {2508.15290},
  primaryclass = {cs},
  publisher = {arXiv},
  doi = {10.48550/arXiv.2508.15290},
  urldate = {2025-12-07},
  archiveprefix = {arXiv}
}

@inproceedings{chenSPANNHighlyefficientBillionscale2021a,
  title = {{{SPANN}}: {{Highly-efficient Billion-scale Approximate Nearest Neighborhood Search}}},
  shorttitle = {{{SPANN}}},
  booktitle = {Advances in {{Neural Information Processing Systems}}},
  author = {Chen, Qi and Zhao, Bing and Wang, Haidong and Li, Mingqin and Liu, Chuanjie and Li, Zengzhong and Yang, Mao and Wang, Jingdong},
  year = 2021,
  volume = {34},
  pages = {5199--5212},
  publisher = {Curran Associates, Inc.},
  urldate = {2025-06-04}
}

@misc{xuScalableDistributedVector2025a,
  title = {Scalable {{Distributed Vector Search}} via {{Accuracy Preserving Index Construction}}},
  author = {Xu, Yuming and Zhang, Qianxi and Chen, Qi and Lu, Baotong and Li, Menghao and Adams, Philip and Li, Mingqin and Li, Zengzhong and Liu, Jing and Li, Cheng and Yang, Fan},
  year = 2025,
  month = dec,
  number = {arXiv:2512.17264},
  eprint = {2512.17264},
  primaryclass = {cs},
  doi = {10.48550/arXiv.2512.17264},
  urldate = {2025-12-24},
  archiveprefix = {arXiv},
  langid = {english}
}

@article{zhuBraveANNRobustApproximate2026,
  title = {{{BraveANN}}: {{Robust Approximate Nearest Neighbor Search}} for {{Billion-Scale Vectors}}},
  shorttitle = {{{BraveANN}}},
  author = {Zhu, Shengkun and Wang, Yiming and Jin, Xin and Zeng, Jinshan and Wang, Sheng and Sun, Yuan and Lai, Yuhui and Peng, Zhiyong},
  year = 2026,
  month = feb,
  journal = {World Wide Web},
  volume = {29},
  number = {1},
  pages = {8},
  issn = {1386-145X, 1573-1413},
  doi = {10.1007/s11280-025-01389-1},
  urldate = {2026-03-23},
  langid = {english}
}

@misc{liuEfficientVectorSearch2025,
  title = {Efficient {{Vector Search}} on {{Disaggregated Memory}} with D-{{HNSW}}},
  author = {Liu, Yi and Fang, Fei and Qian, Chen},
  year = 2025,
  month = may,
  number = {arXiv:2505.11783},
  eprint = {2505.11783},
  primaryclass = {cs},
  publisher = {arXiv},
  doi = {10.48550/arXiv.2505.11783},
  urldate = {2025-06-13},
  archiveprefix = {arXiv}
}

@misc{widmoserSHINEScalableHNSW2025,
  title = {{{SHINE}}: {{A Scalable HNSW Index}} in {{Disaggregated Memory}}},
  shorttitle = {{{SHINE}}},
  author = {Widmoser, Manuel and Kocher, Daniel and Augsten, Nikolaus},
  year = 2025,
  publisher = {arXiv},
  doi = {10.48550/ARXIV.2507.17647},
  urldate = {2026-03-30},
  copyright = {Creative Commons Attribution 4.0 International},
  langid = {english}
}

@inproceedings{liuTigerVectorSupportingVector2025a,
  title = {{{TigerVector}}: {{Supporting Vector Search}} in {{Graph Databases}} for {{Advanced RAGs}}},
  shorttitle = {{{TigerVector}}},
  booktitle = {Companion of the 2025 {{International Conference}} on {{Management}} of {{Data}}},
  author = {Liu, Shige and Zeng, Zhifang and Chen, Li and Ainihaer, Adil and Ramasami, Arun and Chen, Songting and Xu, Yu and Wu, Mingxi and Wang, Jianguo},
  year = 2025,
  month = jun,
  series = {{{SIGMOD}}/{{PODS}} '25},
  pages = {553--565},
  publisher = {Association for Computing Machinery},
  address = {New York, NY, USA},
  doi = {10.1145/3722212.3724456},
  urldate = {2026-01-21},
  isbn = {979-8-4007-1564-8}
}

@misc{niDiskANNEfficientPagebased2023,
  title = {{{DiskANN}}++: {{Efficient Page-based Search}} over {{Isomorphic Mapped Graph Index}} Using {{Query-sensitivity Entry Vertex}}},
  shorttitle = {{{DiskANN}}++},
  author = {Ni, Jiongkang and Xu, Xiaoliang and Wang, Yuxiang and Li, Can and Yao, Jiajie and Xiao, Shihai and Zhang, Xuecang},
  year = 2023,
  month = nov,
  number = {arXiv:2310.00402},
  eprint = {2310.00402},
  primaryclass = {cs},
  publisher = {arXiv},
  doi = {10.48550/arXiv.2310.00402},
  urldate = {2025-12-26},
  archiveprefix = {arXiv}
}

@inproceedings{liuFastVectorSearch2026,
  title = {Fast {{Vector Search}} in {{PostgreSQL}}: {{A Decoupled Approach}}},
  author = {Liu, Jiayi and Zhang, Yunan and Jin, Chenzhe and Gupta, Aditya and Liu, Shige and Wang, Jianguo},
  year = 2026,
  langid = {english},
  booktitle={Conference on Innovative Data Systems Research (CIDR)}
}

\end{document}